\documentclass[letterpaper, oneside, 12pt]{article}
\usepackage{amsmath}
\usepackage{graphicx}%
\usepackage{amsfonts}%
\usepackage{amssymb}
\usepackage{xcolor}
\usepackage{overpic}
\usepackage[noend]{algpseudocode}
\usepackage{subfigure}
\usepackage{rotating}
\usepackage{epstopdf}
\usepackage{url}
\usepackage{comment}
\usepackage{lipsum}
\usepackage{enumitem}
\newlist{steps}{enumerate}{1}
\setlist[steps, 1]{label = Step \arabic*:}
\usepackage{mathtools}
\usepackage{multirow}
\usepackage{pgfplots}
\usepackage{natbib}
\usepackage{pst-solides3d}
\usepackage{booktabs}
\usepackage{tabularray}
\usepackage[section]{placeins}
\usepackage{subcaption}
\usepackage{newfloat}
\usepackage{caption}
\usepackage{lscape}

\usepackage{physics}
\usepackage[normalem]{ulem} 
\usepackage{color}
\usepackage{kotex}
\usepackage{booktabs}
\usepackage{pdfpages}

\usepackage{setspace}
\usepackage[hmargin=1in,vmargin=1in]{geometry}

\usepackage[compact]{titlesec}
\titlespacing{\section}{0pt}{*0.8}{*0.8}
\titlespacing{\subsection}{0pt}{*0.8}{*0.8}
\titlespacing{\subsubsection}{0pt}{*0.8}{*0.8}

\usepackage{A_command}
\usepackage[commandnameprefix=always]{changes}
\usepackage{epstopdf}
\usepackage[flushleft]{threeparttable}

\usepackage{algorithm}
\DeclareCaptionLabelSeparator{space}{\hspace{0.3cm}}
\DeclareCaptionLabelFormat{algolab}{\textbf{ALGORITHM~#2}}
\makeatletter
\newcommand\fs@ruled@notop{\def\@fs@cfont{\bfseries}\let\@fs@capt\floatc@ruled
  \def\@fs@pre{}%
  \def\@fs@post{\kern2pt\hrule\relax}%
  \def\@fs@mid{\kern2pt\hrule\kern2pt}%
  \let\@fs@iftopcapt\iftrue}
\renewcommand\fst@algorithm{\fs@ruled@notop}
\makeatother

\makeatletter
\newcommand*\bigcdot{\mathpalette\bigcdot@{.5}}
\newcommand*\bigcdot@[2]{\mathbin{\vcenter{\hbox{\scalebox{#2}{$\m@th#1\bullet$}}}}}
\makeatother

\newcommand{\expe}{\operatorname{\mathit{E}}}

\newtheorem{theorem}{Theorem}[section]

\newtheorem{corollary}[theorem]{Corollary}

\newcommand{\qed}{\nobreak \ifvmode \relax \else
      \ifdim\lastskip<1.5em \hskip-\lastskip
      \hskip1.5em plus0em minus0.5em \fi \nobreak
      \vrule height0.75em width0.5em depth0.25em\fi}

\date{} 

\title{External Risk Prediction Informed Bayesian Survival Analysis}

\author{Yena Jeon$^{1}$, Yunxiang Huang$^{2}$, Hang J. Kim$^3$, Susan Halabi$^4$, and  Mi-Ok Kim$^{1,*}$ \\
\small
$^{1}$Department of Epidemiology and Biostatistics, University of California, San Francisco, CA, U.S.A. \\
\small
$^{2}$ {Dalian Institute of Chemical Physics, Chinese Academy of Sciences, Liaoning, China} \\
\small
$^{3}$Division of Statistics and Data Science, University of Cincinnati, Cincinnati, OH, U.S.A.\\
\small
$^{4}$Department of Biostatistics and
Bioinformatics, Duke University, Durham, NC, U.S.A.}

\begin{document}

\maketitle
\begin{abstract}
Prognostic factor evaluation and prediction model development are central to precision oncology, enabling patient risk stratification and individualized treatment selection. Unified predictions that synthesize information from existing models are valuable for comprehensive and consistent risk assessment. Many studies also seek to evaluate the incremental value of new biomarkers beyond established prognostic factors. However, such efforts are often constrained by small-to-moderate sample sizes. Motivated by these challenges, we consider Cox regression analysis in settings where individualized risk predictions from existing models are externally available without a transparent or interpretable structure, for example, through online calculators. We develop a Bayesian discretized survival time  inference framework in which individualized predictions from potentially multiple external sources are integrated through a formulation based on Kullback–Leibler divergence, yielding informative priors. The divergence-based formulation serves as a surrogate for the external information likelihood, enabling principled incorporation of individualized predictions without requiring knowledge of the underlying external prediction models. Theoretical results show that the resulting posterior mean estimators are asymptotically more efficient than their internal-only maximum likelihood counterparts. However, using the divergence-based surrogate in place of the unavailable external likelihood renders posterior variance-based inference conservative. We propose a correction to address this overcoverage. We demonstrate the performance of the proposed approach through simulations and an application to prostate cancer trial data.
\end{abstract}
\noindent\emph{Keywords: data integration; external risk informed prior; Halabi's calculator; PREVAIL calculator}.

\section{Introduction} \label{S:intro} 

Prognostic factor evaluation and prediction model development are central to precision oncology, enabling patient risk stratification and individualized treatment selection \citep[e.g.,][]{AlDoughaim2024cancer}. From a clinical perspective, unified predictions that synthesize information from existing models are particularly valuable, as they support comprehensive and consistent risk assessment. At the same time, many studies aim to assess the incremental value of new biomarkers, such as emerging genetic variants, beyond established prognostic factors. These complementary objectives are often constrained by small-to-moderate sample sizes, motivating principled approaches that borrow external information to improve inference and prediction.

An illustrative example is metastatic castration-resistant prostate cancer (mCRPC), an advanced disease with poor prognosis and a median survival of approximately 14 months after the onset of castration resistance \citep{kirby2011characterising}. Among several prognostic models developed for mCRPC, the updated Halabi model \citep{Halabi2014} and the PREVAIL model \citep{PREVAIL1, PREVAIL2} are widely used to predict overall survival. These models rely on established clinical predictors, suggesting potential gains from incorporating genomic biomarkers such as the Decipher classifier. Although originally developed for localized prostate cancer and validated for predicting metastasis and cancer-specific mortality \citep{Den2014GC, Feng2021decipher}, Decipher remains underexplored in mCRPC. The phase III COU-AA-302 trial (\citeauthor{AA302} 2015, NCT00887198), conducted contemporaneously with the trials underlying these models (CALGB-90401: NCT00110214; PREVAIL: NCT01212991), collected genomic data in a subset of participants, providing an opportunity to evaluate its prognostic and predictive value and to improve risk prediction. The COU-AA-302 subset, however, is small, raising the need to incorporate information from the updated Halabi and PREVAIL models to enable a more comprehensive and coherent risk assessment with improved precision.

A substantial body of work has addressed the incorporation of external information involving disparate covariate sets, particularly when such information is available as population-level summary statistics. The empirical likelihood and generalized method of moments frameworks are widely used in this context, representing external information through estimating equations, which serve either as constraints in the empirical likelihood formulation or as additional population moment conditions \citep[e.g.,][for early foundational work]{Qin2000combining, Imbens1994combining}. In survival analysis with Cox regression, these ideas have been extended to combine subgroup-specific survival probabilities or reduced-form regression estimates \citep{huang2016survival, huang2021synthesizing}. More recent work has considered covariate shift, population heterogeneity, and potentially incompatible external information \citep[e.g.,][]{ChenShenQinNing2024Likelihood, ShengSunHuangKim2021Synthesizing, Li2023accommodating}.
These approaches, however, are limited to population-level summaries and require specification of a working model for the external information. 

In contrast, we consider a setting where external information is available at the individual level in the form of predicted risks at specific time points, while the underlying prediction models are unknown. Such predictions may be produced by data-driven or “black-box” algorithms that yield individual risk probabilities without a transparent or interpretable structure, a scenario increasingly common in oncology and precision medicine \citep[e.g.,][]{kourou2015ML}. In other cases, predictions may originate from conventional regression models for which parameter estimates are unavailable, for example, due to proprietary constraints. 

Relatively few methods address this setting. For binary outcomes, \cite{han2022prediction} proposed an empirical likelihood approach using user-specified covariate functions. However, selecting appropriate covariate functions is not straightforward, and the method’s performance is sensitive to this specification. Alternatively, \cite{gu2019synthetic, gu2023synthetic} proposed synthetic-data approaches that generate pseudo-observations consistent with external predictions and combine them with internal data. These methods, however, lack principled guidance for selecting the synthetic sample size and do not explain why efficiency gains plateau beyond a certain point, even when integrating true external information without sampling error. When external information is provided as risk scores rather than full models, \cite{Wang2025KL, Wang2023KL} proposed a penalized partial likelihood approach based on the Kullback--Leibler (KL) divergence \citep{kullback1951information}. The approach penalizes discrepancies between internal model–based and externally provided risk scores. While effective for prediction, it is not designed for inference and relies on ad hoc tuning of the penalty weight parameter, which lacks a clear inferential interpretation.

Motivated by these limitations, we develop a Bayesian Cox proportional hazards regression framework that enables both inference and prediction. Specifically, we introduce a working imputation model to accommodate differences in covariate availability between the internal and external data sources. We then employ KL divergence to align the internal Cox model with the external predictions, thereby inducing an informative prior that incorporates the external individualized risk predictions. Because the divergence-based formulation serves as a surrogate for the unavailable external likelihood, posterior variances are conservative; we develop a correction that restores nominal coverage, yielding valid uncertainty quantification alongside the efficiency gains delivered by the informative prior.

The proposed framework is extended to  address key practical challenges, including potentially heterogeneous baseline hazards across information sources,  uncertainty in external information, and misspecification of the working imputation model.  In particular, the working imputation model can be specified broadly enough to encompass the true data-generating process, without requiring correct specification of the relationships between overlapping and non-overlapping covariates. The proposed approach naturally extends to multiple external sources through source-specific KL divergences. These extensions enhance the generality and flexibility of the proposed approach. 

The rest of the manuscript is organized as follows. For expository simplicity, Section \ref{S:methodology} first considers a single external information source and presents the proposed method along with supporting theoretical justification and Bayesian computation. Section \ref{S:simstudy} reports results from simulation studies. In Section \ref{S:realdata}, we extend the proposed approach to multiple external prediction sources and illustrate its application using the mCRPC data. Concluding remarks are provided in Section \ref{S:conclusion}.

\section{Methodology} \label{S:methodology}

\subsection{Model setup} \label{subs:setup}

Suppose the observed internal study data $\mathcal{D}_n = \{y_i, \Delta_i, \bmx_i, \bmz_i\}_{i=1}^n$ consist of independent and identically distributed copies of $(Y, \Delta, \bmX, \bmZ)$, where $Y = \min(T,C)$ is the observed follow-up time, $\Delta = I(T \leq C)$ is the event indicator, with $T$ and $C$ denoting the event time of interest and the censoring time, respectively, and $(\bmX, \bmZ)$ is a $p$-dimensional covariate vector.  Assuming $T$ and $C$ are independent given $(\bmX, \bmZ)$, the discretized survival-time Cox proportional hazards model specifies the discrete hazard within the study period $[0,\tau]$ as 
\begin{equation} \label{cox}
	\diff \bmLambda(t | \bmx, \bmz; \bmtheta) = 1- \{ 1-\diff \bmLambda(t) \}^{\exp(\bmtheta^T\tilde\bmx)}  ,\; t \in [0,\,\tau],
\end{equation} 
where $\bmLambda(\cdot)$ is an unknown baseline cumulative hazard function, $\tilde\bmx^T =(\bmx^T, \bmz^T)$, and $\bmtheta^T = (\bmtheta_X^T, \bmtheta_Z^T) \in \mathbb{R}^p$ is the vector of covariate effect parameters. In order to address the infinite-dimensional baseline cumulative hazards function $\bmLambda(\cdot)$, we use a piecewise constant baseline cumulative hazard function \citep[see, for example,][]{IbrahimChenSinha2001Bayesian} with increments at certain time points $\mathcal{T} = \{t_j\}_{j=1}^J$ such that $t_1 < t_2 < \cdots < t_J$. From \eqref{cox}, the discretized survival function and its decrement at $t_j$ are respectively given by $S(t_j | \bmx, \bmz; \bmtheta, \bmLambda)
	= \prod_{\{ j': t_{j'} \leq t_j \}} (1 - \lambda_{j'})^{\exp(\bmtheta^T \tilde\bmx)}$ and $f(t_j | \bmx, \bmz; \bmtheta, \bmLambda)
		={\prod_{ \{ j': j' < j \} } (1 - \lambda_{j'})^{\exp(\bmtheta^T \tilde\bmx)} \cdot \qty{1 - (1 - \lambda_j)^{\exp(\bmtheta^T \tilde\bmx)} }}$, for $j=2,\ldots,J$, 
where $\lambda_j$ is the increment of the discretized baseline hazard at $t_j$ and 
$f(t_1 | \bmx, \bmz; \bmtheta, \bmLambda) = 1$ by convention. 

If we only use the internal study data, the baseline cumulative hazard has increments at $\{y_i : \Delta_i=1\}$ and the Bayesian inference for the model parameters $(\bmtheta, \bmLambda)$ is made by the posterior distribution 
\begin{equation} \label{eq:IPDposterior} 
\pi(\bmtheta, \bmLambda | \mathcal{D}_n ) \propto \prod_{i=1}^n  f( y_i | \Delta_i, \bmx_i, \bmz_i; \bmtheta, \bmLambda ) \cdot  \pi( \bmtheta, \bmLambda )
\end{equation} 
where $ f( y_i | \Delta_i, \bmx_i, \bmz_i; \bmtheta, \bmLambda ) = f(y_i | \bmx_i, \bmz_i; \bmtheta, \bmLambda)^{ \Delta_i } 
S(y_i | \bmx_i, \bmz_i; \bmtheta, \bmLambda)^{1 - \Delta_i}$ and $\pi(\bmtheta, \bmLambda)$ denote the prior distribution of the model parameters. 

Suppose auxiliary survival information from an external source is available in the form of {\it individualized}  predictions $P(T \leq t_w^* | \bmX)$ at time points $t_1^* < \dots < t_m^{*}$. For instance, $1-P(T \leq t_w^* | \bmX)$ may represent 3- or 5-year survival probabilities after accounting for individual-level risk factors or predictors included in $\bmX$. The overlapping covariates $\bmX$ may correspond to established risk factors or predictors, while the covariates $\bmZ$, available only in the internal study, may represent a novel risk factor or predictor, such as a new biomarker. Without loss of generality we assume $t_m^{*} \leq \tau$ and reformulate the auxiliary information as the event time information over intervals defined by $0=t_0 < t_1^* < \dots < t_m^{*} < t_{m+1}^* = \infty$ such that $g_W(w|\bmx)=P(T \leq t_w^* | \bmx)-P(T \leq t_{w-1}^* | \bmx)$ for $w=1,\dots,\,m + 1$.  Note $g_W(w|\bmx)$ is the density function for a random variable $W$ with $W=w$ corresponding to $t_{w-1}^* < T \leq t_w^*$. 

\subsection{External prediction informed Bayesian cox regression}
\label{sec:BayesFramework_external}

We use the KL divergence to align the internal Cox model with the externally provided individualized risk predictions $g_W(w|\bmx)$ and to induce an informative prior that reflects the information contained in these external individualized predictions. Suppose the conditional density of the new candidate risk factors given the established risk factors $f( \bmz | \bmx; \bmGamma)$ is known up to the unknown parameter vector $\bmGamma$. Then, the joint density of $( y_i, \bmz_i )$'s under the internal study model is written by 
$\prod_{i=1}^{n} f( y_i | \Delta_i, \bmx_i, \bmz_i; \bmtheta, \bmLambda )$ $f( \bmz_i | \bmx_i; \bmGamma )$ and the prior $\pi(\bmtheta, \bmLambda, \bmGamma)$ is specified for a Bayesian analysis. Under this model, the probability density function of $W$ is
\begin{align} 
    f_W(w | \bmx; \bmtheta, \bmLambda, \bmGamma)
    & = \int \left\{ S(t_{w-1}^* | \bmx, \bmz; \bmtheta, \bmLambda) -  S(t_w^* | \bmx, \bmz; \bmtheta, \bmLambda) \right\} f(\bmz | \bmx;\bmGamma) \dd \bmz \label{eq w2} 
\end{align}
for $w=1,\cdots,m+1$, 
where $S(t_{l}^* | \bmx, \bmz; \bmtheta, \bmLambda)=\prod_{ \{ j:t_j \leq t_{l}^* \} } (1 - \lambda_j)^{\exp( \bmtheta^T \tilde\bmx)}$,  $1\leq l \leq m+1$.

We quantify the discrepancy between the probabilities computed from the model, $f_W(w | \bmx; \\
\bmtheta, \bmLambda, \bmGamma)$, and those based on the external predictions,  $g_W(w|\bmx)$, by the KL divergence as 
\begin{equation} \label{eq w3}
	D_{KL}( g_W( \cdot | \bmx) \| f_W( \cdot | \bmx; \bmtheta, \bmLambda, \bmGamma) )  = \sum_{w=1}^{m+1}  \log \left\{ \frac{g_W(w | \bmx)}{ f_W(w | \bmx; \bmtheta, \bmLambda, \bmGamma) } \right\} g_W(w | \bmx).  
\end{equation}
Interpreting the KL divergence as a measure of information loss incurred by deviation from the externally provided individual-level predictions, we replace the prior distribution $\pi(\bmtheta, \bmLambda, \bmGamma)$ with the following {\it external-prediction-induced prior}:  
\begin{equation} \label{eq:ext-induce-prior}
\pi ( \bmtheta, \bmLambda, \bmGamma | g_W ) = C(g_W) \cdot \exp \left\{ - \sum_{i=1}^{n}  D_{KL} \left( g_W(\cdot|\bmx_i) \| f_W( \cdot | \bmx_i; \bmtheta, \bmLambda, \bmGamma)  \right) \right\} \pi(\bmtheta, \bmLambda, \bmGamma), 
\end{equation}
where  $C(g_W)$ is a normalizing constant such that 
$$
C(g_W)^{-1} = \int \exp \left\{ - \sum_{i=1}^{n}  D_{KL} \left( g_W(\cdot|\bmx_i) \| f_W( \cdot | \bmx_i;  \bmtheta, \bmLambda, \bmGamma)  \right) \right\} \pi(\bmtheta, \bmLambda, \bmGamma) \ \diff (\bmtheta, \bmLambda, \bmGamma).
$$ 
Then, instead of (\ref{eq:IPDposterior}), the Bayesian inference follows from the posterior distribution, 
\begin{equation} \label{E:combined-lik}
    \begin{split}
        \pi(\bmtheta, \bmLambda, \bmGamma | \mathcal{D}_n, g_W) \propto &{} \prod_{i=1}^n f( y_i | \Delta_i, \bmx_i, \bmz_i; \bmtheta, \bmLambda ) f(\bmz_i | \bmx_i; \bmGamma) \cdot \pi ( \bmtheta, \bmLambda, \bmGamma | g_W )\,.
    \end{split}
\end{equation}

Note that the KL divergence-based loss arises as the pointwise limit of the log-likelihood function up to an additive constant, providing a natural surrogate for the external information likelihood within a Bayesian framework. Suppose that we observe independent copies of $W$ for a given $\bmx$ from the conditional distribution $g_W(w | \bmx)$. As the sample size grows, the sample mean of $\log \{f_W(w | \bmx; \bmtheta, \bmLambda, \bmGamma) \} $ converges pointwise to $\expe[\log \{f_W(w | \bmx; \bmtheta, \bmLambda, \bmGamma) \} ]$, which can be written as a constant minus the KL divergence from $g_W(w | \bmx)$ to $f_W(w | \bmx; \bmtheta, \bmLambda, \bmGamma)$. See, for example, Section 5.5 of \cite{vanderVaart1998Asymptotic}, for more details. 

This use of a likelihood surrogate contrasts the proposed approach with fully Bayesian approaches. When the external information is available in the form of population level summaries parameterized by a finite-dimensional vector $\boldsymbol{\beta}$, fully Bayesian methods directly leverage the relationship between $\bmtheta$ and $\boldsymbol{\beta}$ through $\int f(y | \bmx,\bmz;\bmtheta) f(\bmz|\bmx) \diff \bmz = f(y | \bmx; \boldsymbol{\beta})$, thereby incorporating information on $\boldsymbol{\beta}$ via a prior distribution \citep[e.g.,][]{cheng2019informing, boonstra2020bayesian}. However, when the external information is given in the form of $g_W(w | \bmx)$, it is not straightforward to construct an analogous fully Bayesian formulation. The theoretical implications of the use of the surrogate are discussed in Section~\ref{subs:asymptotics} below. 

The proposed approach can also be viewed as improving internal-study analysis in a manner analogous to synthetic data methods, but without explicitly generating synthetic datasets. Specifically, the KL divergence–based term corresponds to the limiting form of a synthetic-data likelihood arising under existing synthetic data approaches \citep[e.g.,][]{gu2019synthetic}, but normalized by the sample size. See Web Appendix A for details. This correspondence provides theoretical insight into the empirical observation of \cite{gu2023synthetic} that efficiency gains from synthetic data plateau beyond a certain synthetic sample size, even when the external information reflects the true underlying model without sampling errors. 

\subsection{Asymptotic properties} \label{subs:asymptotics}

We now consider the large sample properties of the proposed approach. Let $\bmSigma_{\textrm{PH}}$ denote the asymptotic joint covariance function of the maximum partial likelihood estimator and the Breslow estimate of the baseline hazard function $\bmLambda$ under the internal-only Cox proportional hazards model. Suppose that the posterior distribution of $\bmGamma$  based on the internal-only approach is root-$n$ consistent and  asymptotically normal with variance $\bmSigma_{\Gamma}$. Let $(\bar\bmtheta, \bar\bmLambda, \bar\bmGamma)$ denote the posterior mean under the combined posterior distribution in \eqref{E:combined-lik}. The following theorem establishes the asymptotic properties of the posterior mean and a Bernstein–von Mises type result for the posterior distribution. Due to the infinite dimensionality of $\bmLambda$, the Bernstein–von Mises type result only holds in weak topology; the proof is given in Web Appendix B.

\begin{theorem} \label{theo asy} 
	Suppose that Conditions C1--C8 in Web Appendix B.1 hold. If the external prediction model is generated from the same underlying models, $f(t \mid \bmx, \bmz; \bmtheta, \bmLambda)$ and $f(\bmz \mid \bmx; \bmGamma)$, as the internal study, then 
    $n^{1/2}(\bar\bmtheta- \bmtheta_0, \bar\bmLambda- \bmLambda_0, \bar\bmGamma- \bmGamma_0)$
converges weakly to a Gaussian process with mean zero and covariance function $\bmSigma_1 = (\bmV^{-1}+\bmA)^{-1} \bmV^{-1} (\bmV^{-1}+\bmA)^{-1} $, where
\begin{equation*}
\bmV = \begin{pmatrix}
    \bmSigma_{\textrm{PH}} & 0 \\
    0 & \bmSigma_{\bmGamma}
	\end{pmatrix} ,\; \text{ and } \; 
    \bmA = \expe \qty{ \pdv[2]{D_{KL}( g_W(W|\bmX) \, \| \,f_W(W| \bmX ;\bmtheta_0, \bmLambda_0,  \bmGamma_0))}{(\bmtheta, \bmLambda, \bmGamma)} }.
\end{equation*}
The posterior distribution of $n^{1/2}(\bmtheta-\bar\bmtheta, \bmLambda-\bar\bmLambda, \bmGamma- \bar\bmGamma)$ converges weakly to a Gaussian process with mean zero and covariance function $ \bmSigma_{2} = (\bmV^{-1}+\bmA)^{-1}$.
\end{theorem}
The operator $\bmA$ in Theorem~\ref{theo asy} captures the curvature of the KL divergence at the true parameter values and thus quantifies the informativeness of the external predictions. By the convexity of the KL divergence, $\bmA$ is positive-definite. When $\bmA$ is large in the positive-definite sense, the external predictions carry substantial information and the combined estimator achieves meaningful variance reduction relative to the internal-only approach.

The asymptotic variance of the posterior mean differs from the posterior variance $\bmSigma_2$.  Specifically, $\bmSigma_2 - \bmSigma_1$ is positive definite, making inference based on the posterior variance alone conservative. This discrepancy arises because the KL divergence–based prior serves as a likelihood surrogate rather than a true likelihood, a phenomenon that is well recognized in the Gibbs posterior and quasi-posterior literature \citep[e.g.,][]{Chernozhukov2003, Bissiri2016}. Fortunately, there is a simple relationship linking the two covariance functions through the information operator $\bmV^{-1}$ of the full likelihood under the internal-only Cox model, summarized in the following corollary. 
\begin{corollary}
	Under the conditions of Theorem \ref{theo asy},     $\bmSigma_{1} =  \bmSigma_{2} \bmV^{-1} \bmSigma_{2}$.
\end{corollary}
As $\bmSigma_{2}$ can be estimated from the combined-data posterior variance and $\bmV^{-1}$ from the internal-only data, the corollary provides a straightforward estimator of $\bmSigma_{1}$. All empirical results in later sections are based on this corrected variance estimate. For more details on the information operator for the proportional hazards model, $\bmV^{-1}$, we refer to page 384 of \cite{Kosorok2008Introduction}. 

Thus far, we have restricted the increments of the baseline hazard to occur only at the set of observed event times, $\mathcal{T}_0 = \{y_i: \Delta_i = 1,\, i=1,\dots,\, n\}$. Under the internal-only proportional hazards model, the full likelihood is maximized when the baseline hazard has increments exclusively at $\mathcal{T}_0 $. In the proposed framework, which incorporates predictions at time points $\{t_w^*\}_{w=1}^m$ with some $t_w^* \notin \mathcal{T}_0$, it is natural to ask whether this property still holds. We have theoretical results that it is sufficient in most cases to restrict the set of increment locations to $\mathcal{T}_0$ for computation. See Theorem 2 in Web Appendix B.3 for details. 

\subsection{Practical considerations and extension} \label{sec:model_detail}

We extend the framework to address several key practical considerations, thereby yielding a more general and flexible inferential procedure. The proposed framework assumes that the conditional distribution of the new risk factors is correctly specified as $f(\bmz \mid \bmx; \bmGamma)$ up to an unknown parameter vector $\bmGamma$, and that the conditional distributions $f(t \mid \bmx, \bmz; \bmtheta, \bmLambda)$ and $f(\bmz \mid \bmx; \bmGamma)$ are shared between the internal and external studies, while allowing the distribution of $\bmX$ to differ across data sources.

These assumptions are comparable to those required by many existing approaches for incorporating population-level summaries, in that they replace assumptions on the imputation model $f(\bmz \mid \bmx; \bmGamma)$ with assumptions on the joint covariate distribution of $(\bmX, \bmZ)$. For example, some approaches require the distribution of $(\bmX, \bmZ)$ to be identical across studies \citep[e.g.,][]{huang2016survival,zhang2020generalized}, whereas others rely on the availability of a large external dataset sampled from the external covariate distribution \citep{kundu2019generalized}, or employ semiparametric methods that use internal-study covariate data to address covariate shift in the external study \citep[e.g.,][]{ShengSunHuangKim2021Synthesizing}. 
Despite these variations, they all require some form of cross-study comparability in covariate distributions. 

In the following, we relax the assumptions of the proposed method to improve robustness and applicability.
We also accommodate uncertainty in the external predictions. 
Although external predictions are typically based on a large sample relative to the internal study, their estimation errors may not be negligible. Extension to multiple external prediction 
sources is deferred to Section \ref{sec:morethanone}.  

\subsubsection{Allowing heterogeneity in the baseline hazard} \label{sec:model_heterogeneity}

We retain the homogeneity assumption that the parametric component (the covariate effects $\bmtheta$) is shared across internal and external populations—otherwise, borrowing information about $\bmtheta$ would not be meaningful. In contrast, the baseline hazard may not be exactly the same because of differences in the study populations or study conduct. We allow adjustments to the baseline hazard by introducing a scale-shift parameter $\nu$ and replace (\ref{eq w2}) with
\begin{equation} \label{eq w2-adjusted} f_W(w | \bmx; \bmtheta, \nu, \bmLambda, \bmGamma)=\int \left\{ S(t_{w-1}^* | \bmx, \bmz; \bmtheta, \nu, \bmLambda) -  S(t_w^* | \bmx, \bmz; \bmtheta, \nu, \bmLambda) \right\} f(\bmz | \bmx;\bmGamma) \dd \bmz 
\end{equation}
where $S(t_{w}^* | \bmx, \bmz; \bmtheta, \nu, \bmLambda)=\prod_{ \{ j:t_j \leq t_{w}^* \} } (1 - \lambda_j)^{\exp( \bmtheta^T \tilde\bmx + \nu)}$ for $w=1,\cdots,m+1$.

We estimate $\nu$ by  empirically imposing $m+1$ constraints, following the Bayesian generalized method of moments \citep{Yin2009Bayesian}: 
$f(\nu | \mathcal{X}_n, g_W; \bmtheta, \bmLambda, \bmGamma ) \propto \det|\bmS_q|^{-1/2} \exp \qty( -\frac{n}{2} \bar{\bmq}^{T} \bmS_q^{-1} \bar{\bmq} )$, 
where $\mathcal{X}_n = \{ \bmx_i\}_{i=1}^n$, $\bar{\bmq}=\sum_{i=1}^{n} \bmq_{i} / n$ and 
$\bmS_q = \sum_{i=1}^{n} \left( \bmq_{i} - \bar{\bmq} \right)^{\otimes 2} / n$ with 
$$ \bmq_i=\begin{pmatrix} 
	f_W(w=1 | \bmx_i, \bmtheta, \nu, \bmLambda, \bmGamma)  \vspace{-3mm}\\
	\vdots \vspace{-4mm}\\
	f_W(w = m+1 | \bmx_i, \bmtheta, \nu, \bmLambda, \bmGamma)
\end{pmatrix} - \begin{pmatrix} 
	\hat{g}(1,\bmx_i) \vspace{-3mm}\\
	\vspace{-4mm}\vdots \\
	\hat{g}(m+1,\bmx_i)
\end{pmatrix}\,.$$
The solution of the moment equation minimizes the discrepancy between $f_W( \cdot | \bmx_i; \bmtheta, \nu, \bmLambda, \bmGamma)$ and $g_W( \cdot | \bmx_i)$.
The {\it external-prediction-induced prior} in (\ref{eq:ext-induce-prior}) is updated to 
\begin{equation} \label{eq:ext-induce-prior 2}
    \pi ( \bmtheta, \nu, \bmLambda, \bmGamma | g_W ) 
   \, \propto \exp \left\{ - \sum_{i=1}^{n}  D_{KL} \left( g_W(\cdot|\bmx_i) \| f_W( \cdot | \bmx_i; \bmtheta, \nu, \bmLambda, \bmGamma)  \right) \right\} \pi(\bmtheta, \nu, \bmLambda, \bmGamma). 
\end{equation} 
 
\subsubsection{Modeling of the conditional distribution of $\bmZ$} \label{sec:model_Z}

This modeling is auxiliary to the Cox model inference and may be approached as prediction modeling. This motivates 
specifying a flexible working model 
such that $f(\bmz | \bmx; \bmGamma)$ is rich enough to nest the unknown true conditional distribution of $\bmZ= (Z_1, \ldots, Z_{p_z})$, 
for example, by including interactions or nonlinear terms,  $\eta\{\mathit{E}(Z_k|\bmx)\}=\gamma_{k0} + \sum_{l=1}^{p_{\gamma_k}} \gamma_{kl} \psi_{kl}(\bmx)$, for some link function $\eta\{\cdot\}$. 
In practice, $p_{\gamma_k}$ may become too large to be reliably estimated, in which case we use penalized estimation. For instance, in our real study application, we assume LASSO-type constraints via Laplace priors (see details in Web Appendix C.1). Accordingly, $\bmGamma$ in the external-information-induced prior in \eqref{eq:ext-induce-prior 2} denotes the set of all parameters pertaining to the working model of $\bmZ$. We examine the robustness of the proposed penalized estimation in the simulation studies in Section \ref{Subs:Scenario2} below.

\subsubsection{Allowing sampling variability in external risk predictions} 
\label{sec:sampling_variability}

We now consider a case in which the external prediction information $g_W(\cdot|\bmx)$ is {\it not} exactly known and is estimated with sampling error. Suppose that, for each $\bmx_i$ and $w=1,\ldots,m+1$, we have the external risk prediction estimate $\hat{g}(w,\bmx_i)$ and its variance estimate $\widehat{V}(\hat{g}(w,\bmx_i))$. 
For these sample proportions, we assume the Dirichlet distribution $(\hat{g}(1,\bmx_i),\ldots,\hat{g}(m+1,\bmx_i)) \sim \text{Dir}( g_{W}(1|\bmx_i)/\upsilon,$ 
$\ldots,g_{W}(m+1|\bmx_i)/\upsilon)$
where the moments are approximated as $E\{ \hat{g}(w,\bmx_i) \} = g_W(w|\bmx_i)$ and $V \{ \hat{g}(w,\bmx_i) \} = E[ \widehat{V}\{ \hat{g}(w,\bmx_i) \} ]$ 
$ = \{ \upsilon/ ( \upsilon+1 ) \} g_W(w|\bmx_i) \{ 1 - g_W(w|\bmx_i)\}$ for some $\upsilon > 0$.

Let $\hat{g}_W = \{ \hat{g}(w,\bmx_i): w=1,\ldots,m+1, i=1,\ldots,n \}$ and assume a flat prior for $g_W$, deriving  $\pi( g_W | \hat{g}_W ) \propto f(\hat{g}_W | g_W)$.   
Then, the  {\it external-prediction-induced prior} in \eqref{eq:ext-induce-prior 2} is replaced by $\pi ( \bmtheta, \nu, \bmLambda, \bmGamma | \hat{g}_W ) = \int \pi ( \bmtheta, \nu, \bmLambda, \bmGamma | g_W )\pi( g_W | \hat{g}_W ) \diff g_W$, resulting in the posterior distribution 
\begin{align}\label{eq combi lik2}
      &{} \pi(\bmtheta, \nu, \bmLambda, \bmGamma, g_W | \mathcal{D}_n, \hat{g}_W) \\
        &{}
        \propto \prod_{i=1}^n f( y_i | \Delta_i, \bmx_i, \bmz_i; \bmtheta, \bmLambda ) f(\bmz_i | \bmx_i; \bmGamma)
        \cdot f(\nu | \mathcal{X}_n; g_W, \bmtheta, \bmLambda, \bmGamma ) \pi(\bmtheta, \nu, \bmLambda, \bmGamma | g_W ) \pi(g_W | \hat{g}_W). \nonumber
\end{align}

Some prediction models provide confidence intervals in practice \citep[e.g., the updated Halabi model,][]{Halabi2014}, from which variances can be derived under the assumption of normal theory–based confidence intervals.
Given the sample proportions and their variance estimates 
$\hat{g}(w,\bmx_i)$ and $\widehat{V}(\hat{g}(w,\bmx_i))$, we approximate the Dirichlet distribution, $f(\hat{g}_W | g_W)$, using the following moment equations: 
$$ \bmm_i (\hat{g}_W) = \begin{pmatrix}
     \hat{g}(1,\bmx_i) - g_W(1|\bmx_i) \vspace{-3mm}\\
\vdots \vspace{-4mm}\\
     \hat{g}(m+1,\bmx_i) - g_W(m+1|\bmx_i) \\
     \widehat{V}\{ \hat{g}(1,\bmx_i) \} - \frac{\upsilon}{\upsilon+1}  g_W(1|\bmx_i)\left\{ 1-g_W(1|\bmx_i) \right\} \vspace{-3mm}\\
\vdots \vspace{-4mm}\\
     \widehat{V}\{ \hat{g}(m+1,\bmx_i) \} - \frac{\upsilon}{\upsilon+1} g_W(m+1|\bmx_i) \left\{ 1-g_W(m+1|\bmx_i) \right\}
 \end{pmatrix}  = \bmzero. $$

\subsubsection{Bayesian computation and inference}
\label{sec:Bayescomputation}

The posterior distribution in \eqref{eq combi lik2} encompasses all extensions accommodating heterogeneous baseline hazard, a working imputation model for $Z$, and sampling error of external risk predictions. We use the Metropolis-within-Gibbs for computation. Appropriate weakly informative priors are assigned to the model parameters, and the cumulative baseline hazard is modeled using a Beta process prior. See Web Appendix C.2 for details of prior specifications and Markov Chain Monte Carlo computation for the model parameters.

\section{Simulation Studies} \label{S:simstudy}
We conducted simulation studies to evaluate the performance of the proposed method under two scenarios, generating 300 datasets for each scenario with an internal sample size of $n=100$. Survival times were simulated from a Weibull distribution with a hazard function 
$$ h(t|\bmX,Z;\bmtheta,\rho, \kappa, \nu) = \kappa^{-1} e^\nu \rho t^{1/{\kappa}-1}\exp(\theta_1 X_1 +\theta_2 X_2 + \theta_3 Z) \,,$$
where the simulation parameters were set as $\bmtheta = (\theta_1, \theta_2, \theta_3) = (0.5, -0.5, 0.5)$, $\rho = \sqrt{2}$, and $\kappa = 0.5$. For the internal study, we set the scale-shift parameter $\nu$ as zero; in the external study, it was specified according to each simulation scenario. The data-generating mechanisms for $\bmX$ and $Z$ also varied across scenarios, as described below. The censoring times were generated from an exponential distribution with mean 0.3 and truncated at an administrative time $\tau_c$, chosen to yield an overall censoring rate of 25\%. For the external survival prediction model, we assumed individualized predictions were provided at two pre-specified time points such that 
$g_W(w|\bmx)=P(T \leq t_w^* | \bmX)-P(T \leq t_{w-1}^* | \bmX)$ was given for $w=1,2,3$ with $t_3^\ast = \tau_c$.

\subsection{Scenario 1: External prediction with certainty and homogeneous data distribution}

This setting is one in which the double empirical likelihood (DEL) approach \citep{huang2016survival} is applicable. Unlike the proposed method, which directly incorporates individual-level prediction information, the DEL approach instead incorporates subgroup-level summaries derived from such information.   We examined the relative efficiency of the two methods. 

We generated $X_1$ and $X_2$ independently from $\text{Uniform}(-0.5, 0.5)$ for both the internal and external populations. The non-overlapping covariate $Z$ was generated from a linear model, $Z = 0.5X_1 + 0.5X_2 + \epsilon$, where $\epsilon \sim N(0,0.1)$ and  $N(\cdot,\cdot)$ denotes normal distributions. We assumed a common baseline hazard distribution across internal and external studies by setting $\nu=0$. The true external information at each threshold $t_w^\ast$ was computed under the assumed Cox model via Monte Carlo integration and treated as known without uncertainty. 

For the DEL approach, we assume that the external source provides subgroup-specific means defined by dichotomizing $X_1$ and $X_2$, respectively. The approach was applied separately under two configurations: using subgroup means defined by ${X_1 \leq 0}$ vs. ${X_1 > 0}$ (DEL-$X_1$), and by ${X_2 \leq 0}$ vs. ${X_2 > 0}$ (DEL-$X_2$). When four subgroups defined by all combinations of $X_1 \leq 0$ vs. $X_1 > 0$ and $X_2 \leq 0$ vs. $X_2 > 0$ were considered, the computation was less stable, and this configuration was not pursued further. We also conducted the internal-only Bayesian analysis as a reference.

The results are summarized in terms of empirical bias, relative efficiency (RE; defined as the ratio of the mean squared error of the internal-only Bayesian approach to that of the proposed method), and coverage probability in Table~\ref{tab:simul 1}. While the REs of $\theta_3$ estimates (for the non-overlapping covariate $Z$) are comparable across methods, those of $\theta_1$ and $\theta_2$ estimates differ substantially, with the proposed {\it external prediction informed} (EPI) approach consistently yielding the largest values. As expected, the DEL methods improve efficiency only for the parameter corresponding to the covariate used for subgrouping, and the improvement is modest compared with that of the proposed method. This highlights the loss of information incurred when individualized information is collapsed into subgroup-level summaries. Bias is small, and the coverage probabilities are close to the nominal 95\% level across all methods. Table D.1 in Web Appendix provides a detailed summary of the results from the imputation model.

\begin{table}[ht]
    \centering
    \caption{Summary statistics for regression coefficients $\hat\bmtheta$ under Scenario 1, which assumes error-free external predictions with certainty and a common covariate and baseline hazard distribution between the internal and external populations. We report bias, relative efficiency (RE), and coverage probability (CP) for the internal-only approach, the double empirical likelihood method (DEL), and the proposed external prediction informed Cox (EPI) model. DEL-$X_1$ and DEL-$X_2$ incorporate auxiliary subgroup survival means defined by dichotomizing $X_1$ and $X_2$, respectively.} \label{tab:simul 1}
    \begin{tabular}{l rrr @{\hskip 6pt} rrr @{\hskip 6pt} rrr}
    \hline
     & \multicolumn{3}{c}{$\theta_1$} & \multicolumn{3}{c}{$\theta_2$} & \multicolumn{3}{c}{$\theta_3$} \\
    \cmidrule(lr){2-4} \cmidrule(lr){5-7} \cmidrule(lr){8-10}
    Methods             & bias & RE & CP & bias & RE & CP & bias & RE & CP \\
    \midrule
    Internal-only     & 0.068 &     1.000 & 0.940 & -0.039 &     1.000 & 0.947 & -0.016 &     1.000 & 0.933  \\
    DEL-$X_1$           & 0.034 & 1.455 & 0.950 & -0.028 & 0.996 & 0.953 & -0.025 & 0.997 & 0.937  \\
    DEL-$X_2$           & 0.060 &     1.000 & 0.950 & -0.022 & 1.283 & 0.940 & -0.025 & 1.005 & 0.940  \\
    EPI               & 0.042 & 2.111 & 0.953 & -0.020 & 2.096 & 0.967 & -0.014 & 0.995 & 0.957  \\
    \bottomrule
    \end{tabular}
\end{table}  

\subsection{Scenario 2: External prediction with sampling errors and heterogeneous data distribution} \label{Subs:Scenario2}

We evaluated the proposed extensions designed to improve the generalizability and flexibility of the proposed framework in this setting. While the data-generating mechanism for the internal study was the same as in Scenario I, Scenario II introduced population heterogeneity by allowing the external population to have a different baseline hazard distribution with $\nu = 0.5$. The covariates for the external study were generated as $X_1 \sim \text{Beta}(1,3)-0.5$ and $X_2 \sim N(0,1)$, while $Z$ was generated in the same way as the internal study. We also allowed sampling variability in the external prediction by generating the external prediction information from an external sample of size $n_{ext}=5,000$. We refer to Web Appendix D.1 for details of data generation and prediction modeling with variance estimates.

To examine the full extension of the proposed method, we let a working model of $Z$ include an interaction term, quadratic terms, and additional variables $X_3$ and $X_4$:
    $$ Z 
    = \gamma_0 + \gamma_1 X_1 + \gamma_2 X_2 + \gamma_3 X_1X_2 + \gamma_4X_1^2 + \gamma_5X_2^2 + \gamma_6 X_3 + \gamma_7 X_4 + \epsilon, \quad \epsilon \sim N(0, 0.1), $$ 
where $X_3 \sim N(0, 0.25)$ and $X_4 \sim \text{Bernoulli}(0.4)$. The variables $X_3$ and $X_4$ are not part of the true conditional distribution of $Z$; including them in the working model tests whether the LASSO-type penalty shrinks their coefficients toward zero while preserving the contributions of $X_1$ and $X_2$.

In order to assess the impacts of the different extensions, we considered three variants of the proposed approach. 
The first, EPI, is the fully extended model that accommodates heterogeneous baseline hazards, sampling errors in the external predictions, and a flexible working model for $Z$. This working model was estimated under LASSO-type constraints implemented through a Laplace prior (see Web Appendix C.1 for details). The second, EPI-trueZ, assumes the true imputation model for $Z$ while still accommodating heterogeneous baseline hazards and incorporating sampling errors in the external predictions. The third, EPI-oracle, is an oracle version, in which the baseline hazard shift is known, the imputation model for $Z$ is correctly specified, and the external prediction information is error-free. 

The results are summarized in Table~\ref{tab:simul 2} in terms of empirical bias, RE, and coverage probability. The REs of all parameter estimates are similar across the three variants of the proposed approach. As expected, the oracle version EPI-oracle yields the largest REs, but the fully extended model EPI—which relies on the least amount of correct specification—performs comparably, with REs of 2.194 and 2.211 for $\theta_1$ and $\theta_2$ against 2.280 and 2.287 for EPI-oracle. Interestingly, the REs under EPI are slightly larger than those under EPI-trueZ, although the differences are modest. The gaps may reflect Monte Carlo variability across 300 replications or may reflect a bias–variance trade-off induced by the additional modeling flexibility incorporated for the working impuation model for $Z$ in EPI.  Overall, all three variants are more efficient than the internal-only approach, exhibit small bias, and attain coverage probabilities close to the nominal 95\% level. Web Appendix Table D.2 provides a summary of the imputation model estimation results. 

\begin{table}[ht]
    \centering
    \caption{Summary statistics for regression coefficients $\hat\bmtheta$ under Scenario 2, which assumes external predictions subject to sampling error and baseline hazard that differ between the internal and external populations. We report bias, relative efficiency (RE), and coverage probability (CP) for the internal-only approach and three variants of the proposed approach: EPI, the fully extended version of the proposed external prediction informed Cox model that accomodates heterogeneous baseline hazards, sampling errors in the external predictions, and a working model for $Z$; EPI-trueZ, which assumes the true imputation model for $Z$ while allowing heterogeneous baseline hazards and sampling errors in the external predictions; and EPI-oracle, an oracle version assuming a known baseline hazard shift, a correctly specified imputation model, and error-free external predictions.} \label{tab:simul 2}
    \begin{tabular}{lrrr @{\hskip 6pt} rrr @{\hskip 6pt} rrr}
    \hline
     & \multicolumn{3}{c}{$\theta_1$} & \multicolumn{3}{c}{$\theta_2$} & \multicolumn{3}{c}{$\theta_3$} \\
    \cmidrule(lr){2-4} \cmidrule(lr){5-7} \cmidrule(lr){8-10}
    Methods     & bias & RE & CP & bias & RE & CP & bias & RE & CP \\
    \midrule
    Internal-only      & 0.068 &     1.000 & 0.940 & -0.039 &     1.000 & 0.947 & -0.016 &     1.000 & 0.933  \\
    EPI   & 0.038 & 2.194 & 0.953 & -0.004 & 2.211 & 0.967 & -0.008 & 0.976 & 0.943  \\ 
    EPI-trueZ                & 0.036 & 2.138 & 0.960 & -0.015 & 2.080 & 0.970 &  0.000 & 0.935 & 0.957  \\
    EPI-oracle         & 0.035 & 2.280 & 0.963 & -0.007 & 2.287 & 0.970 & -0.024 & 1.034 & 0.960  \\
    \bottomrule
    \end{tabular}
\end{table}

In this scenario, we also evaluated predictive performance. For each simulated dataset, we generated an independent validation dataset of size $n_{\text{test}}=1{,}000$ from the internal data-generating process and computed the root mean squared prediction error (RMSPE) at time $t$ as  
 $  \text{RMSPE}(t) = \sqrt{ \sum_{i=1}^{n_{\textrm{test}}} \left\{ {S}(t|\bmx_i,z_i;\hat\bmtheta,\hat\Lambda) - S(t|\bmx_i,z_i;\bmtheta_0,\rho_0,\kappa_0) \right\}^2 / n_{\textrm{test}} } \,,$
where ${S}(t|\bmx_i,z_i;\hat\bmtheta,\hat\Lambda)$ is the survival probability predicted from the fitted discretized Cox model using  using posterior means $\hat\bmtheta$ and $\hat\Lambda$, and $S(t|\bmx_i,z_i;\bmtheta_0,\rho_0,\kappa_0)$ is the true survival probability under the data-generating Weibull model.

\begin{figure} 
    \centering
    \includegraphics[width=1\linewidth]{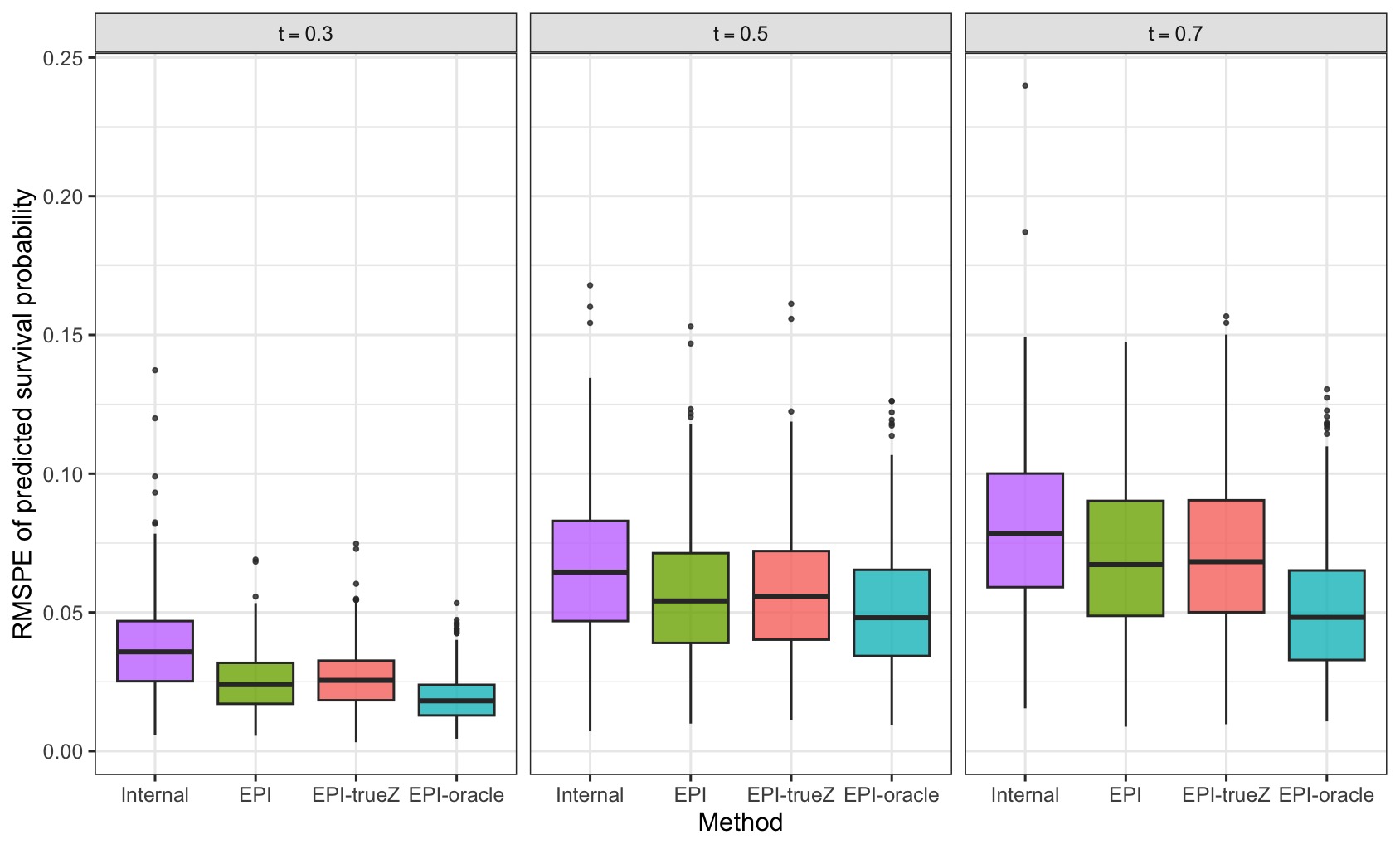}
    \caption{Distribution of the root mean squared prediction error (RMSPE) of survival probabilities at $t=0.3, 0.5,$ and $0.7$ comparing the internal-only approach and three variants of the proposed approach based on a validation set of size $n_\text{test}=1000$. EPI denotes the fully extended external prediction-informed Cox model allowing heterogeneous baseline hazard, sampling errors in the external predictions, and a working model for $Z$. EPI-trueZ assumes true imputation model for $Z$ while allowing heterogeneous baseline hazard and sampling errors in the external predictions. EPI-oracle denotes a gold-standard version assuming known baseline shifts, a correctly specified imputation, and error-free external prediction information.}
    \label{fig:pred_rmse}
\end{figure}

Figure~\ref{fig:pred_rmse} presents the distribution of RMSPE$(t)$ of the survival probabilities at time points $t=0.3,0.5$, and $0.7$ based on the validation dataset. All three EPI variants outperform the internal-only approach across time points, with EPI-oracle achieving the lowest RMSPE as expected. EPI and EPI-trueZ yield closely matching RMSPE distributions at each time point, indicating that introducing a flexible working model for $Z$ does not compromise predictive performance. The RMSPE increases at later time points for all methods, likely reflecting greater uncertainty in survival probabilities as the effective sample size at risk decreases. Overall, these results suggest that leveraging external information through the proposed EPI framework improves predictive performance relative to using internal data alone.

\section{Metastatic Castration-Resistant Prostatee Cancer Overall Survival Prediction} \label{S:realdata}

As discussed in the Introduction, our goal is to enhance the updated Halabi model \citep{Halabi2014} and PREVAIL model \citep{PREVAIL1, PREVAIL2} by incorporating genomic information from the promising Decipher genomic classifier, using data from the COU-AA-302 randomized phase III trial \citep[][NCT00887198]{AA302}. Although COU-AA-302 enrolled 1,088 patients, Decipher measurements are available for only 98 patients. This limited subset is insufficient for reliably evaluating the incremental prognostic value of Decipher by jointly incorporating the Decipher score and the established clinical predictors used in the updated Halabi and PREVAIL models. Because both external models provide individualized survival predictions through online calculators, we treat the COU-AA-302 subset as the internal study of interest and apply the proposed method to augment Cox regression risk modeling using these external predictions.

Several considerations support the proposed augmentation. First, the randomized trials underlying the updated Halabi and PREVAIL models show strong temporal and clinical comparability with the COU-AA-302 trial. These trials were conducted over a similar time frame (approximately 2005--2012) and used broadly similar inclusion and exclusion criteria. Across studies, participants were required to have an ECOG performance status below 2 and serum testosterone below 50 ng/dL, indicating a consistent mCRPC disease state. In addition, the control groups received similar standards of care. Details of the randomized trials with cohort characteristics are provided in Web Appendix Table E.1. 

Integrating predictions from these two external sources introduces several challenges. First, the updated Halabi and PREVAIL models do not use identical predictor sets, and the time points at which survival predictions are available do not fully overlap. Specifically, the updated Halabi online calculator provides predicted survival probabilities with 95\% confidence intervals at 18, 24, 30, 36, and 48 months, whereas the PREVAIL online calculator reports only point predictions at 1, 2, 3, and 5 years without accompanying measures of uncertainty. In addition, the COU-AA-302 subset includes patients from both the placebo and treatment arms, introducing additional heterogeneity relative to the populations used to develop the external models.

We address these challenges as follows. First, we compute survival predictions under the assumption that all patients received placebo, thereby aligning with the risk prediction settings of both external models and providing a suitable baseline for borrowing. Differences in predictors and non-overlapping prediction time points are handled, using the extension of the proposed approach to multiple external sources described below in Section \ref{sec:morethanone}. Because the PREVAIL development trial was approximately seventeen times larger than the internal study, uncertainty in its predictions is treated as negligible. In contrast, for the updated Halabi model, we use Dirichlet distributions to model individualized prediction distributions and account for sampling uncertainty, following Section \ref{sec:sampling_variability}.  

\subsection{Extension to multiple external prediction models and real study computation} \label{sec:morethanone} 

The proposed framework naturally accommodates external predictions from multiple sources. When $K$ multiple external models are available, as in our real-data analysis with $K=2$, we partition the internal study covariate vector into ``shared'' and ``internal-only'' depends on which external model is being considered. Following the notations in Section \ref{subs:setup}, we write $\tilde{\bmx} = (\bmx^{(k)}, \bmz^{(k)})$ for $k=1,\ldots,K$, where $\bmx^{(k)}$ denotes the shared covariates also used by the $k$-th external model and $\bmz^{(k)}$ the non-overlapping remainder.

For each external source $k$, let $W^{(k)}$ denote the corresonding individualized prediction variable with externally provided distributions $g_{W^{(k)}}( w^{(k)} | \bmx^{(k)})$. Following,  Section~\ref{subs:setup}, we construct corresponding model-implied distribution $f_{W^{(k)}}( w^{(k)} | \bmx^{(k)}; \bmtheta, \bmLambda, \bmGamma)$ and formulate external source specific KL divergences as $D_{KL}( g_{W^{(k)}}( \cdot | \bmx^{(k)}) \| f_{W^{(k)}}( \cdot | \bmx^{(k)}; \bmtheta, \bmLambda, \bmGamma) )$ accordingly. This formulation accommodates external prediction defined on different covariate sets and at different time points across sources. When the sampling variability in the external predictions is negligible, the {\it external-prediction-induced prior} generalizes to $\pi ( \bmtheta, \bmLambda, \nu, \bmGamma | g_{W^{(1)}},\ldots,g_{W^{(K)}}) \propto$ $ \exp \{ - \sum_{k=1}^{K} \sum_{i=1}^{n}$  $D_{KL} ( g_{W^{(k)}}(\cdot|\bmx_i^{(k)}) \|  f_{W^{(k)}}( \cdot | \bmx_i^{(k)}; \bmtheta, \nu, \bmLambda, \bmGamma)  ) \}$ $\pi(\bmtheta, \nu, \bmLambda, \bmGamma)$. 

\subsection{Real study analysis results} \label{sec:realstudy}

We applied the fully extended proposed approach to the real mCRPC study analysis, accommodating the two external prediction sources and other practical challenges. Prior specification and computational details follow those used in the simulations and are summarized in Web Appendix C. Interestingly, the original regression estimates from the updated Halabi and PREVAIL models are available, which we include under the third and fourth columns of Table~\ref{tab:realdata Cox}. The table also reports the results of the internal-only Bayesian Cox regression analysis estimates and the EPI estimates. 

\begin{table}[ht]
    \centering
    \caption{Estimated regression coefficients (Est.) and standard errors (SE) of the Cox models for the prostate cancer study. EPI is the fully extended external prediction informed Cox model that allows heterogeneous baseline hazard, sampling errors in the external predictions, and a working model for $Z$. $\nu^H$ and $\nu^P$ are the parameters accounting for the baseline hazard shift of Halabi and Prevail model, respectively. Columns labeled Halabi and PREVAIL report the regression coefficient estimates copied from the original studies.} \label{tab:realdata Cox}
    \begin{tabular}{l c cc @{\hskip 6pt} ccc}
    \hline
        & Internal-only  & EPI  & Halabi & PREVAIL  \\
    \cmidrule(lr){2-3} \cmidrule(lr){4-5} 
    Variables             & Est. (SE) & Est. (SE) & Est. (SE) & Est. (SE) \\
    \midrule
    log(PSA)              &  0.153 (0.103) &  0.031 (0.029) & {\color{black} 0.009 (0.006)} & {\color{black} 0.090 (0.016)} \\
    LDH$\geq$ULN          & -0.012 (0.492) &  0.029 (0.084) & {\color{black} 0.146 (0.039)} & {\color{black} 0.174 (0.046)} \\
    Albumin               &  0.210 (0.442) &  0.186 (0.322) & {\color{black}-0.051 (0.029)} & {\color{black}-0.149 (0.065)} \\
    Hemoglobin            &  0.002 (0.140) & -0.025 (0.049) & {\color{black}-0.027 (0.014)} & {\color{black}-0.071 (0.018)} \\
    ALP$\geq$ULN          &  0.366 (0.360) &  0.107 (0.102) & {\color{black} 0.064 (0.029)} & {\color{black} 0.102 (0.047)} \\
    Bone metastases       &  0.452 (0.353) &  0.551 (0.380) &           --   & {\color{black} 0.167 (0.044)} \\
    ECOG                  & -0.959 (0.399) & -0.158 (0.102) & {\color{black} 0.134 (0.035)} & \\
    Pain score$\geq$2     &  0.364 (0.328) &  0.115 (0.125) & & {\color{black} 0.114 (0.040)} \\
    Time from diagnosis   & -0.041 (0.031) & -0.012 (0.013) & & {\color{black}-0.001 (0.000)} \\
    Decipher score        &  0.017 (0.070) &  0.013 (0.022) & & \\
    Treatment group       &  0.053 (0.293) & -0.107 (0.320) & & \\
    \midrule
    $\nu^H$  &  &  0.201 (0.236) &  &  \\
    $\nu^P$  &  & -1.019 (0.227) &  &  \\
    \bottomrule
    \end{tabular}
    \vspace{2em}
    \parbox{\linewidth}
    {\footnotesize
    Abbreviations: PSA, prostate-specific antigen; LDH, lactate dehydrogenase; ULN, upper limit of normal; ALP, alkaline phosphatase; ECOG, Eastern Cooperative Oncology Group performance status.
    }
\end{table}

The proposed approach, EPI, yields uniformly smaller standard errors than the internal-only analysis, reflecting substantial efficiency gains from borrowing external information. The estimates are overall similar between the updated Halabi and PREVAIL models, whereas the estimates obtained from the internal-only analysis for the overlapping prognostic factors differ substantially from those of either external model. In some cases, the directions of the effects are even reversed. These discrepancies are consistent with instability of the internal-only estimates given the small sample size.

The proposed method addresses such instability. EPI shrinks unstable internal estimates toward the more reliable external estimates, yielding results that may be interpreted as data-driven weighted averages of the internal-only and externally provided estimates. Such results would be expected if coefficient estimates from the external models were directly incorporated under the assumed knowledge of their respective modeling structures. Instead, the proposed method achieves a similar effect by incorporating individualized predictions, without requiring access to the underlying external models.

We highlight the results for two variables: ECOG and the treatment group. ECOG is a measure of patients’ functional status, where higher values indicate poorer performance and are therefore associated with lower survival. In the internal-only analysis, the estimated coefficient for ECOG is $-0.959$ (SE = 0.399), suggesting that worse functional status is associated with longer survival, which is clinically implausible. However, with the external predictions incorporated EPI reports an attenuated, non-significant effect. On the other hand, COU-AA-302 was a positive trial in which the treatment arm demonstrated improved survival relative to the placebo group. However, the internal-only analysis yielded a positive coefficient estimate (0.053), whereas the EPI estimate was $-0.107$, which is directionally consistent with the overall trial result, although not statistically significant in this subset.
EPI mitigates the instability arising from the small convenience subset by borrowing strength from external information, leading to more stable and coherent inference. 

The scale-shift parameters $\nu^H$ and $\nu^P$ quantify baseline-hazard heterogeneity between the two external models and the internal study, respectively. Whereas the estimated value of $\nu^H$ is small and statistically non-significant, the estimated value of $\nu^P$ is $-1.019$, suggesting substantial heterogeneity in the baseline hazards between the internal cohort and the PREVAIL development population. Motivated by concerns regarding potential inconsistency in the PREVAIL model, we conducted an additional sensitivity analysis using only the Halabi model predictions. As expected, the efficiency gain is less pronounced. Overall, the results are consistent with those of the primary analysis (see Web Appendix Table E.3 and Table E.4). 
Estimates from the working imputation model for $Z$ are auxiliary and are summarized in Web Appendix E.2.

\section{Concluding Remarks} \label{S:conclusion}

In this paper, we propose a Bayesian framework for integrating external individual-level prediction information from potentially multiple sources into Cox regression analysis. The proposed method fills an important gap in the existing literature, which has largely focused on incorporating population-level summary information and lacks principled methods for integrating individualized predictions without assuming knowledge of the underlying prediction models. Within the proposed Bayesian framework, the KL divergence–based integration provides a surrogate likelihood for the external prediction information and induces an informative prior.

The proposed approach is general and flexible, as it does not require the internal and external studies to share the same baseline hazards or covariate distributions, nor does it require correct specification of the working imputation model. However, the method does not address potentially inconsistent external information and the resulting risk of bias. This limitation could be addressed through incorporation strategies that adapt to the degree of commensurability between the internal and external data, which we leave for future work.

\section*{Acknowledgements}
The authors are grateful to Drs. Eric Small and Julian Hong for providing the deidentified data used in the empirical application of this paper.

\section*{Supplementary Materials}
Additional methodological details, simulation results, and other supporting materials are provided in the Web Appendix.

\section*{Funding} 
This research was supported by the Patient-Centered Outcomes Research Institute (PCORI) Award under Grant ME-2020C3-21269.

\bibliography{A_reference}

\end{document}


\def\spacingset#1{\renewcommand{\baselinestretch}%
		{#1}\small\normalsize} \spacingset{1}
	
	
		\title{Supplementary Material for  ``External Risk Prediction Informed Bayesian Survival Analysis'' \\

        \vspace{1em}
        
        {\large Yena Jeon$^{1}$, Yunxiang Huang$^{2}$, Hang J. Kim$^3$, Susan Halabi$^4$, and  Mi-Ok Kim$^{1,*}$} \\
        
        \vspace{1em}
        
        {\small 
        $^{1}$Department of Epidemiology and Biostatistics, University of California, San Francisco, CA, U.S.A. 
        \vspace{0.5em}
        
        $^{2}$Dalian Institute of Chemical Physics, Chinese Academy of Sciences, Liaoning, China 
        \vspace{0.5em}
        
        $^{3}$Division of Statistics and Data Science, University of Cincinnati, Cincinnati, OH, U.S.A.
        \vspace{0.5em}
        
        $^{4}$Department of Biostatistics and Bioinformatics, Duke University, Durham, NC, U.S.A. 
        \vspace{0.5em}

        $^*$email: miok.kim@ucsf.edu
        
        } }

		\date{}
		\maketitle
	
	\spacingset{1.5}


\renewcommand{\thesection}{\Alph{section}} 

\section{Connection with Synthetic Data Approaches}

The proposed approach can be viewed as improving internal study analysis like synthetic data methods, but without explicitly generating synthetic datasets. Following existing synthetic data approaches \citep[e.g.,][]{gu2019synthetic}, suppose the parametric form of $P(\bmZ | \bmX, W)$ is known, and the model parameters are estimated from the internal study data. One could construct a synthetic dataset by first sampling $\{{\bmx}^\ast_j\}$ via resampling from the observed internal covariate data, then drawing ${w}^\ast_j$ from $g_{W}(\cdot | {\bmx}^\ast_j)$, and subsequently drawing $\bmz^\ast_j$ from the estimated $P(\bmZ | \bmX, W)$ with $\bmX={\bmx}^\ast_j$ and $W={w}^\ast_j$. If such synthetic datasets are generated repeatedly and in sufficiently large numbers, this procedure isapproximately equivalent to, for each observed $\bmx_i$, drawing $({w}^\ast_{ij}, \bmz^\ast_{ij})$ for $j=1,\cdots, N$, where $N$ is a sufficiently large number. 
On the other hand from Equations (4)--(6) from the main text, we have 
\begin{equation*}
    \begin{split}
    \log \pi(\bmtheta, \bmLambda, \bmGamma | \mathcal{D}_n, g_W) \propto 
        &{} \sum_{i=1}^n \log \{f( y_i | \Delta_i, \bmx_i, \bmz_i; \bmtheta, \bmLambda ) f(\bmz_i | \bmx_i; \bmGamma)\} \\
        &{} + \sum_{i=1}^{n} \sum_{w \in \Omega} \log f_{W}(w | \bmx_i; \bmtheta, \bmLambda, \bmGamma) g_W(w | \bmx_i) + \log \pi(\bmtheta, \bmLambda, \bmGamma)\,,
    \end{split}
\end{equation*}
where $\Omega$ denotes the support of $W$. Given $\{({w}^\ast_{ij}, \bmz^\ast_{ij})\}$ for each $\bmx_{i}$, we further note from Equation (3) in the main text
$$
\sum_{w \in \Omega} \log f_{W}(w | \bmx_i; \bmtheta, \bmLambda, \bmGamma) g_W(w | \bmx_i) \approx \dfrac{1}{N} \sum_{j=1}^{N} \log \{S(t^*_{{w}^\ast_{ij}-1} | \bmx_i, \bmz^\ast_{ij}; \bmtheta, \bmLambda) -  S(t^*_{{w}^\ast_{ij}} | \bmx_i, \bmz^\ast_{ij}; \bmtheta, \bmLambda) \}.$$ 
This connection between the KL divergence–based term and the limiting form of a normalized synthetic-data likelihood indicates that the proposed approach recovers the same information as repeatedly generating synthetic datasets with $N \to \infty$, but does so analytically and without generating auxiliary synthetic datasets. It also provides theoretical insight into the empirical observation in \cite{gu2023synthetic} that efficiency gains from synthetic data plateau beyond a certain synthetic sample size, even when the external information reflects the underlying truth. 

\vspace{3em} 
\section{Proofs of Theorems}

\subsection{Proof of Theorem 1}

The following conditions are required:
\begin{itemize}[itemsep=1pt]
    \item[(C1)] $T$ is independent of $C$ when $\widetilde{\bmx}$ is given.
    \item[(C2)] $\pr(T > \tau | \widetilde{\bmx} = 0) > \kappa$ and $\pr(C = \tau | \widetilde{\bmx}) > \kappa$ almost surely for some $\kappa >0$.
    \item[(C3)] The underlying cumulative baseline hazard function is absolutely continuous and strictly increasing on $[0,\tau]$.
    \item[(C4)] The covariance matrix of $Z$ is of full rank.
    \item[(C5)] $Z$ is bounded almost surely.
    \item[(C6)] $\pi(\bmtheta)$ and $\pi(\bmGamma)$ are continuous at $\bmtheta_0$ and $\bmGamma_0$, respectively.
    \item[(C7)] The scale parameter of the beta process prior, denoted by $c(\cdot)$, satisfies $0 <\inf_{t \in [0,\tau]} c(t) < \sup_{t \in [0,\tau]} c(t) < \infty$
    \item[(C8)] The posterior distribution of $\Gamma$ based on the IPD only approach is root-$n$ consistent and asymptotically normal with variance $\bmSigma_{\Gamma}$.
\end{itemize}
Conditions (C1)--(C4) are common under the proportional hazards model.
See, for example, Section 3 of \cite{Kim2006Bernsteinvon}. 
Condition (C5) is for technical purposes, which can be replaced by $Z$, which is sub-Gaussian, with verifying the existing results in \cite{Kim2006Bernsteinvon}.
Conditions (C6) and (C7) are required for the priors.
When the beta process prior to the cumulative baseline hazard function satisfies Condition (C7), the conditions for the prior in Theorem 3.3 of \cite{Kim2006Bernsteinvon} are always satisfied.
Condition (C8) requires that the behavior of the IPD only approach for $\Gamma$ is good.
See, for example, Chapter 10 of \cite{vanderVaart1998Asymptotic} for more details.

Let $(\widehat{\bmtheta}, \widehat{\bmLambda})$ and $(\widetilde{\bmtheta}, \widetilde{\bmLambda})$ respectively denote the maximum likelihood estimator associated with the external prediction-informed likelihood function and the posterior mode under the proposed approach.
To simplify notation, we omit the superscript of $\bmLambda_{\tau}$ throughout the proof.
We divide the proof into two parts.
We first establish the weak convergence of the maximum likelihood estimator to a Gaussian process.
Second, we prove a Bernstein--von Mises type result for the posterior distribution under the proposed approach.

\subsubsection{Property of the maximum likelihood estimator} \label{sec MLE}
Let 
\begin{equation*}
    L_{\textrm{PH}}(\bmtheta, \bmLambda) = \prod_{i=1}^{n} \{f(Y_i | \bmz_i,\bmx_i; \bmtheta, \bmLambda_{\tau})\}^{ \Delta_i } 
	\{S(Y_i | \bmz_i,\bmx_i; \bmtheta, \bmLambda) \}^{1 - \Delta_i}
\end{equation*}
denote the full likelihood function based on the IPD observations under the proportional hazards model, whose logarithm is denoted by $\ell_{\textrm{PH}}(\bmtheta, \bmLambda)$.
Recall that $\mathcal{T}_0 = \{t_j
\}_{j=1}^{J_0} = \{Y_i: \Delta_i = 1\}$.
Then we have
\begin{equation} \label{eq ipdlik}
    L_{\textrm{PH}}(\bmtheta, \bmLambda)
    = \prod_{i=1}^n \prod_{t_j \in \mathcal{T}_0: t_j < Y_i} (1 - \lambda_{j})^{\exp(\bmtheta^{\top} \widetilde{\bmx}_i)} \qty{1 - (1 - \dd{\bm\Lambda}(Y_i))^{\exp(\bmtheta^{\top} \widetilde{\bmx}_i)}}^{\Delta_i}  .
\end{equation}

The IPD-only maximum likelihood estimator that maximizes $ L_{\textrm{PH}}(\bmtheta, \bmLambda)$ is denoted by $(\widehat{\bmtheta}_{\textrm{IPD}}, \widehat{\bmLambda}_{\textrm{IPD}})$.
Let $\widehat{\bmtheta}_{\textrm{PL}}$ be the maximum partial likelihood estimator which maximizes the logarithm of the partial likelihood
\begin{equation*}
    \ell_{PL}(\bmtheta) = 
\sum_{i=1}^n \Delta_i \left[
\bmtheta^\top \widetilde{\bmx}_i 
- \log \left\{ \sum_{j \geq i} \exp (\bmtheta^\top \widetilde{\bmx}_j) \right\} \right].
\end{equation*}
Let $\widehat{\bmLambda}_{\textrm{PL}}$ be the Breslow’s estimator of the baseline hazard function, which is given by
\begin{equation*}
    \widehat{\bmLambda}_{\textrm{PL}} (t)
    = \int_0^t \frac{\dd{N}(s)}{\sum_{j: Y_j \geq s} \exp(\widehat{\bmtheta}_{\textrm{PL}}^\top \widetilde{\bmx}_j) },\; t \in [0,\tau],\;\;
    N(\cdot) = \sum_{i=1}^n \Delta_i I(Y_i \leq \cdot).
\end{equation*}
Our aim is to prove $\|\widehat{\bmtheta}_{\textrm{IPD}} - \widehat{\bmtheta}_{\textrm{PL}}\| = O_p(n^{-1})$ and 
$|\dd\widehat{\bmLambda}_{\textrm{IPD}}(t) - \dd\widehat{\bmLambda}_{\textrm{PL}}(t)| = o_p(n^{-3/2})$ uniformly over $t \in \{t_j\}_{j=1}^{J_0} = \{Y_i: \Delta_i = 1\}$.

We first consider $\widehat{\bmLambda}_{\textrm{IPD}}(\cdot)$.
Here, our task is to prove $|\dd\widehat{\bmLambda}_{\textrm{IPD}}(t) - \dd\widehat{\bmLambda}_{\textrm{PL}}(t)| = o_p(n^{-3/2})$ holds when $\|\widehat{\bmtheta}_{\textrm{IPD}} - \widehat{\bmtheta}_{\textrm{PL}}\| = o_p(n^{-1/2})$.
Let $u_i = (1 - \lambda_{Y_i})^{\exp(\bmtheta^\top \widetilde{\bmx}_i)}$ for $i \in \{k: \Delta_k = 1\}$. 
Then we can rewrite the IPD-only full likelihood in \eqref{eq ipdlik} as
\begin{equation*}
    \begin{split}
        L_{\textrm{PH}}(\bmtheta, u) 
            & = \prod_{i=1}^n \prod_{t_j \in \mathcal{T}_0: t_j < Y_i} u_j^{\exp(\bmtheta^\top \widetilde{\bmx}_i) / \exp(\bmtheta^\top \widetilde{\bmx}_j)} (1 - u_i)^{\Delta_i} \\
            & = \prod_{t_j \in \mathcal{T}_0} (1 - u_j)
            \prod_{i: Y_i > t_j} u_j^{\exp(\bmtheta^\top \widetilde{\bmx}_i) / \exp(\bmtheta^\top \widetilde{\bmx}_j)},
    \end{split}
\end{equation*}
whose logarithm is given by
\begin{equation}
    \ell_{\textrm{PH}}(\bmtheta, u)
    = \sum_{t_j \in \mathcal{T}_0} \log(1 - u_j) + 
      \sum_{t_j \in \mathcal{T}_0} e^{-\bmtheta^\top \widetilde{\bmx}_j} \qty( \sum_{i: Y_i > t_j} e^{\bmtheta^\top \widetilde{\bmx}_i} ) \log u_j .
\end{equation}
By taking the partial derivative with respect to $u_j$ and setting it to zero, we have
\begin{equation*}
    -\frac{1}{1- u_j} + e^{-\bmtheta^\top \widetilde{\bmx}_j} \qty( \sum_{i: Y_i > t_j} e^{\bmtheta^\top \widetilde{\bmx}_i} ) \frac{1}{u_j} = 0,\; j=1,\dots, J_0.
\end{equation*}
Consequently, when $\bmtheta$ is given, the optimal $u_j$ is given by
\begin{equation*}
    u_j(\bmtheta) = 1 - \frac{e^{\bmtheta^\top \widetilde{\bmx}_j}}{\sum_{i: Y_i \geq t_j} e^{\bmtheta^\top \widetilde{\bmx}_i}} ,\; j=1,\dots, J_0.
\end{equation*}
And the corresponding $\lambda_j$ satisfies
\begin{equation} \label{eq lambdaj 1}
    \lambda_j(\bmtheta)
    = 1 - \qty{ 1 - \frac{e^{\bmtheta^\top \widetilde{\bmx}_j}}{\sum_{i: Y_i \geq t_j} e^{\bmtheta^\top \widetilde{\bmx}_i}}  }^{\exp(-\bmtheta^\top \widetilde{\bmx}_j)} .
\end{equation}
Let $\Theta_0 = \{\bmtheta: \|\bmtheta - \bmtheta_0\| \leq \kappa_{\bmtheta}\}$ for some $\kappa_{\bmtheta} > 0$.
By Condition (C2), we can conclude that, uniformly over $\bmtheta \in \Theta_0$, 
\begin{equation*}
    \frac{1}{\sum_{i: Y_i = \tau} e^{\bmtheta^\top \widetilde{\bmx}_i}} = O(n^{-1})
\end{equation*}
holds almost surely.
By using Taylor expansion up to the second order for $\lambda_j(\cdot)$ in \eqref{eq lambdaj 1}, we can further conclude that
\begin{equation*}
    \lambda_j(\bmtheta) =  \frac{1}{\sum_{i: Y_i \geq t_j} e^{\bmtheta^\top \widetilde{\bmx}_i}} + O(n^{-2})
    =  \dd\widehat{\bmLambda}_{\textrm{PL}}(t_j) + n^{-1} O(\|\bmtheta - \widehat{\bmtheta}_{\textrm{PL}}\|) + O(n^{-2})
\end{equation*}
uniformly over $j = 1,\dots J_0$ and $\bmtheta \in \Theta_0$ almost surely, which completes this step of the proof.

We next consider $\widehat{\bmtheta}_{\textrm{IPD}}$.
Let $\rho(\bmtheta) = \max_{\bmLambda} \ell_{\textrm{PH}}(\bmtheta, \bmLambda) =
\max_{u}\ell_{\textrm{PH}}(\bmtheta, u)$ denote the profile likelihood function of $\bmtheta$.
Then we have
\begin{equation*}
    \begin{split}
    \dv{\rho(\bmtheta)}{\bmtheta}
    ={}& \sum_{t_j \in \mathcal{T}_0} \log\qty[1 - \frac{e^{\bmtheta^\top \widetilde{\bmx}_j}}{\sum_{i: Y_i \geq t_j} e^{\bmtheta^\top \widetilde{\bmx}_i}} ] \sum_{i: Y_i > t_j} (\widetilde{\bmx}_i - \widetilde{\bmx}_j)e^{\bmtheta^\top (\widetilde{\bmx}_i - \widetilde{\bmx}_j)}  \\
    ={}& -\sum_{t_j \in \mathcal{T}_0} \frac{e^{\bmtheta^\top \widetilde{\bmx}_j}}{\sum_{i: Y_i \geq t_j} e^{\bmtheta^\top \widetilde{\bmx}_i}} \sum_{i: Y_i > t_j} (\widetilde{\bmx}_i - \widetilde{\bmx}_j)e^{\bmtheta^\top (\widetilde{\bmx}_i - \widetilde{\bmx}_j)} + O(n^{-1}) \\
    ={}& -\sum_{t_j \in \mathcal{T}_0} \frac{\sum_{i: Y_i \geq t_j} (\widetilde{\bmx}_i - \widetilde{\bmx}_j)e^{\bmtheta^\top \widetilde{\bmx}_i}}{\sum_{i: Y_i \geq t_j} e^{\bmtheta^\top \widetilde{\bmx}_i}} + O(n^{-1}) \\
    ={}&  \dv{\ell_{PL}(\bmtheta)}{\bmtheta}  + O(n^{-1})
    \end{split}
\end{equation*}
uniformly over $\bmtheta \in \Theta$ almost surely.
By Taylor expansion of $\dv*{\ell_{PL}(\bmtheta)}{\bmtheta} $ at $\bmtheta = \widehat{\bmtheta}_{\textrm{PL}}$, we further have
\begin{equation*}
     0 = \dv{\rho(\widehat{\bmtheta}_{\textrm{IPD}})}{\bmtheta}
     = \dv{\ell_{PL}(\widehat{\bmtheta}_{\textrm{IPD}})}{\bmtheta}  + O(n^{-1})
     = {\dv[2]{\ell_{PL}(\widehat{\bmtheta}_{\textrm{PL}})}{\bmtheta}}  (\widehat{\bmtheta}_{\textrm{IPD}} - \widehat{\bmtheta}_{\textrm{PL}})
     + o(\|\widehat{\bmtheta}_{\textrm{IPD}} - \widehat{\bmtheta}_{\textrm{PL}}\|) +  O(n^{-1})
\end{equation*}
almost surely.
Then $\|\widehat{\bmtheta}_{\textrm{IPD}} - \widehat{\bmtheta}_{\textrm{PL}}\| = O_p(n^{-1})$ follows immediately.

We next consider the asymptotic property of the maximum likelihood estimator under the proposed approach.
Let $\textrm{BV}_{\tau}$ denote the family of functions of bounded variation on $[0,\tau]$.
Let $\bmSigma_{\textrm{PH}}$ be the asymptotic covariance function for $n^{1/2}(\widehat{\bmtheta}_{\textrm{PL}} - \bmtheta_0, \widehat{\bmLambda}_{\textrm{PL}} - \bmLambda_0)$.
Then the operator associated with $\bmSigma_{\textrm{PH}}$ is continuously invertible and onto on $\mathbb{R}^{p+1} \times \textrm{BV}_{\tau}$ equipped with norm
\begin{equation*}
    \|h\|_{\mathcal{H}} = \qty[ \|h_1\|_2^2 + \int_0^\tau [h_2(t)]^2 \dd{\bmLambda_0(t)}]^{1/2},\;\; \textrm{for}\; h = (h_1,h_2) \in \mathbb{R}^p \times \textrm{BV}_{\tau},
\end{equation*}
where $\|\cdot\|_2$ denotes the Euclidean norm of a vector.
See Page 384 of \cite{Kosorok2008Introduction} for more details.
The components of the information operator $\sigma_{\textrm{PH}} = \bmSigma_{\textrm{PH}}^{-1}$ are given by
\begin{equation*}
    (\sigma_{\textrm{PH}})_{\bmtheta\bmtheta} h_1
    = \int_0^{\tau} \expe[ \widetilde{\bmx} \widetilde{\bmx}^\top I(Y \geq s) e^{\bmtheta^\top \widetilde{\bmx}}] \dd{\bmLambda_0(s)} h_1,
\end{equation*}
\begin{equation*}
    (\sigma_{\textrm{PH}})_{\bmtheta\bmLambda} h_2
    = \int_0^{\tau} \expe[ \widetilde{\bmx} I(Y \geq s) e^{\bmtheta^\top \widetilde{\bmx}}] h_2(s) \dd{\bmLambda_0(s)} ,
\end{equation*}
\begin{equation*}
    (\sigma_{\textrm{PH}})_{\bmLambda\bmtheta} h_1(\cdot) 
    = \expe[ \widetilde{\bmx}^\top I(Y \geq \cdot) e^{\bmtheta^\top \widetilde{\bmx}}] h_1,
\end{equation*}
\begin{equation*}
    (\sigma_{\textrm{PH}})_{\bmLambda\bmLambda} h_2(\cdot) 
    = \expe[ I(Y \geq \cdot) e^{\bmtheta^\top \widetilde{\bmx}}] h_2(\cdot).
\end{equation*}

By the asymptotic equivalence between  $(\widehat{\bmtheta}_{\textrm{IPD}}, \widehat{\bmLambda}_{\textrm{IPD}})$ and $(\widehat{\bmtheta}_{\textrm{PL}}, \widehat{\bmLambda}_{\textrm{PL}})$, we also have $\vari\{n^{1/2}(\widehat{\bmtheta}_{\textrm{IPD}} - \bmtheta_0, \widehat{\bmLambda}_{\textrm{IPD}} - \bmLambda_0)\} \to \bmSigma_{\textrm{PH}}$.
Since $(\widehat{\bmtheta}_{\textrm{IPD}}, \widehat{\bmLambda}_{\textrm{IPD}})$ is estimated by maximizing the IPD-only likelihood, we can conclude that
\begin{equation*}
    -\frac{1}{n}\pdv[2]{\ell_{\textrm{PH}}(\widehat{\bmtheta}_{\textrm{IPD}}, \widehat{\bmLambda}_{\textrm{IPD}})}{(\bmtheta, \bmLambda)} - \bmSigma_{\textrm{PH}} = o_p(1).
\end{equation*}
Then by Taylor expansion, we have, for any $(\bmtheta, \bmLambda) \in \mathbb{R}^{p+1} \times \textrm{BV}_{\tau}$
\begin{equation*}
    \begin{split}
        \ell_{\textrm{PH}}(\bmtheta, \bmLambda) ={}& \ell_{\textrm{PH}}(\widehat{\bmtheta}_{\textrm{IPD}}, \widehat{\bmLambda}_{\textrm{IPD}}) 
    + \frac{1}{2}{\pdv[2]{\ell_{\textrm{PH}}(\widehat{\bmtheta}_{\textrm{IPD}}, \widehat{\bmLambda}_{\textrm{IPD}})}{(\bmtheta, \bmLambda)}}
    [\bmtheta - \widehat{\bmtheta}_{\textrm{IPD}}, \bmLambda- \widehat{\bmLambda}_{\textrm{IPD}}]^{\otimes 2} \\
    &{}+ o\{ \|\bmtheta - \widehat{\bmtheta}_{\textrm{IPD}}, \bmLambda- \widehat{\bmLambda}_{\textrm{IPD}}\|_{\mathcal{H}}^2 \}.
    \end{split}
\end{equation*}
The convexity of the Kullback--Leibler divergence implies
\begin{equation*}
    \pdv{D_{\textrm{KL}}(g_W(W | \widetilde{\bmX})\, \| \,f_W(W | \widetilde{\bmX};  \bmtheta_0, \bmLambda_0, \bmGamma_0)}{(\bmtheta, \bmLambda, \bmGamma)} = 0
\end{equation*}
for any $\widetilde{\bmX}$.
Let $\ell(\bmtheta, \bmLambda, \bmGamma)$ denote the log-likelihood function of the proposed approach at a certain point.
Then, by the second-order Taylor expansion, we further have, uniformly over $
\{(\bmtheta, \bmLambda, \bmGamma) : \|\bmtheta - \widehat{\bmtheta}_{\textrm{IPD}}, \bmLambda- \widehat{\bmLambda}_{\textrm{IPD}}\|_{\mathcal{H}}^2 +  (\bmGamma - \widehat{\bmGamma}_{\textrm{IPD}})^2 \leq n^{-1/3}\}$,
\begin{equation*}
    \begin{split}
        &{} \ell(\bmtheta, \bmLambda, \bmGamma) - \ell_{\textrm{PH}}(\widehat{\bmtheta}_{\textrm{IPD}}, \widehat{\bmLambda}_{\textrm{IPD}})  - \ell_{\textrm{IPD}}(\widehat{\bmGamma}_{\textrm{IPD}}) \\
        ={}&  \frac{1}{2}{\pdv[2]{\ell_{\textrm{PH}}(\widehat{\bmtheta}_{\textrm{IPD}}, \widehat{\bmLambda}_{\textrm{IPD}})}{(\bmtheta, \bmLambda)}}  [\bmtheta - \widehat{\bmtheta}_{\textrm{IPD}}, \bmLambda- \widehat{\bmLambda}_{\textrm{IPD}}]^{\otimes 2}
          + \frac{1}{2}{\pdv[2]{\ell_{\textrm{IPD}}(\widehat{\bmGamma}_{\textrm{IPD}})}{\bmGamma}} (\bmGamma - \widehat{\bmGamma}_{\textrm{IPD}})^2 \\
          &{} - \frac{1}{2}{ \pdv[2]{D_{\textrm{KL}}(g_W(W | \widetilde{\bmX})\, \| \,f_W(W | \widetilde{\bmX};  \bmtheta_0, \bmLambda_0, \bmGamma_0)}{(\bmtheta, \bmLambda, \bmGamma)} } [\bmtheta - \bmtheta_0, \bmLambda- \bmLambda_0,\bmGamma - \bmGamma_0]^{\otimes 2} \\
          &{} + o\{ (\bmGamma - \widehat{\bmGamma}_{\textrm{IPD}})^2 + \|\bmtheta - \widehat{\bmtheta}_{\textrm{IPD}}, \bmLambda- \widehat{\bmLambda}_{\textrm{IPD}}\|_{\mathcal{H}}^2 \} \\
          ={}& \frac{1}{2}V^{-1}[\bmtheta - \widehat{\bmtheta}_{\textrm{IPD}}, \bmLambda- \widehat{\bmLambda}_{\textrm{IPD}},\bmGamma - \widehat{\bmGamma}_{\textrm{IPD}}]^{\otimes 2} + \frac{1}{2}A[\bmtheta - \bmtheta_0, \bmLambda- \bmLambda_0,\bmGamma - \bmGamma_0]^{\otimes 2} \\
          &{} + o_p\{ (\bmGamma - \widehat{\bmGamma}_{\textrm{IPD}})^2 + \|\bmtheta - \widehat{\bmtheta}_{\textrm{IPD}}, \bmLambda- \widehat{\bmLambda}_{\textrm{IPD}}\|_{\mathcal{H}}^2 \}
    \end{split}
\end{equation*}
where the last equation results from the law of large numbers.
The components of $A$ correspondingly have the form
\begin{equation*}
    A_{\bmLambda\bmLambda}(s,t) = \sum_{j=1}^m \sum_{j'=1}^m A_{\bmLambda\bmLambda,jj'} \delta(t_j^* - t)\delta(t_{j'}^* - s),\, \textrm{for } (t,s) \in [0,\tau]
\end{equation*}
and
\begin{equation*}
    A_{\bmLambda\bmtheta}(t) = \sum_{j=1}^m  A_{\bmLambda\bmtheta,j} \delta(t_j^* - t),\, \textrm{for } t \in [0,\tau],
\end{equation*}
where $A_{jj'} \in \mathbb{R}$, $A_{\bmLambda\bmtheta}(\cdot) \in \mathbb{R}^p$, and $\delta(\cdot)$ is the Dirac delta function.
Consequently, the log-likelihood function $ \ell(\bmtheta, \bmLambda, \bmGamma)$ is maximized at
\begin{equation} \label{eq mle}
    (\widehat{\bmtheta}, \widehat{ \bmLambda}, \widehat{\bmGamma}) = (V^{-1} + A)^{-1} \{V^{-1}(\widehat{\bmtheta}_{\textrm{IPD}}, \widehat{\bmLambda}_{\textrm{IPD}}, \widehat{\bmGamma}_{\textrm{IPD}}) + A (\bmtheta_0, \bmLambda_0, \bmGamma_0)\} + o_p(n^{-1/2}) .
\end{equation}
Therefore, $n^{1/2}(\widehat{\bmtheta} - \bmtheta_0, \widehat{ \bmLambda} - \bmLambda_0, \widehat{\bmGamma} -  \bmGamma_0)$ converges weakly to a Gaussian process with mean zero.
Additionally, taking the covariance on both sides yields
\begin{equation*}
    \vari\{n^{1/2}(\widehat{\bmtheta}, \widehat{ \bmLambda}, \widehat{\bmGamma})\} \to  (V^{-1} + A)^{-1} V^{-1}   (V^{-1} + A)^{-1},
\end{equation*}
which completes the proof.

\subsubsection{Posterior inference}

Let $\widehat{\expe}_{\textrm{IPD}}(\cdot)$ and $\widehat{\expe}(\cdot)$ denote the posterior expectations of a certain function of $(\bmtheta, \bmLambda, \bmGamma)$ under the IPD-only and the proposed approaches, respectively.
We divide the proof of the asymptotic properties of the posterior distribution into three steps.
First, we summarize some existing asymptotic results for the posterior distribution of the IPD-only approach.
Second, we prove the Bernstein–von Mises type result for the proposed approach by directly verifying the characteristic function of the posterior distribution.
In the last step, we prove $ \widehat{\expe}[n^{1/2}(|\bmtheta - \widehat{\bmtheta}|, |\bmLambda(\cdot) -  \widehat{\bmLambda}(\cdot)|, |\bmGamma - \widehat{\bmGamma}|)]$ is bounded almost surely.
The consistency of the posterior mean then follows immediately using the dominated convergence theorem.

Step 1.  By Theorem 3.3 of \cite{Kim2006Bernsteinvon}, the posterior distribution of 
$n^{1/2}(\bmtheta - \widehat{\bmtheta}_{\textrm{PL}}, \bmLambda -  \widehat{\bmLambda}_{\textrm{PL}})$ based on the IPD observations under the proportional hazards model converges weakly to a zero-mean Gaussian process with the covariance function $\bmSigma_{\textrm{PH}}$.
Note that both the IPD-only likelihood on $(\bmtheta, \bmLambda, \bmGamma) $ and the priors are separable in $(\bmtheta, \bmLambda)$ and $\bmGamma$.
Therefore, the posterior distribution of $(\bmtheta, \bmLambda)$ and $\bmGamma$ are independent.
By the standard result of Bayesian maximum likelihood approach, we can also conclude that $n^{1/2}(\bmGamma - \widehat{\bmGamma}_{\textrm{IPD}})$ converges in distribution to a zero-mean Gaussian distribution with the covariance matrix $\bmSigma_{\bmGamma}$.
See, for example, Chapter 10 of \cite{vanderVaart1998Asymptotic}.
Then it immediately follows that the posterior distribution of 
$n^{1/2}(\bmtheta - \widehat{\bmtheta}_{\textrm{PL}}, \bmLambda -  \widehat{\bmLambda}_{\textrm{PL}}, \bmGamma - \widehat{\bmGamma}_{\textrm{IPD}})$ converges weakly to a zero-mean Gaussian process with the covariance function $\bmV$.

Step 2. 
Define the event
\begin{equation} \label{ev e}
    \mathcal{E}_n = \qty{ \widehat{\expe}_{\textrm{IPD}}\qty[ \exp\qty{ 
    -\sum_{i=1}^n D_{\textrm{KL}}(g_W(W | \widetilde{\bmx}_i)\, \| \,f_W(W | \widetilde{\bmx}_i;  \bmtheta, \bmLambda, \bmGamma)}] \in (0,1] }.
\end{equation}
We first prove the event $\mathcal{E}_n$ holds almost surely as $n\to \infty$.
By the non-negativity of the Kullback--Leibler divergence, it immediately follows that
\begin{equation*}
    \widehat{\expe}_{\textrm{IPD}}\qty[ \exp\qty{ 
    -\sum_{i=1}^n D_{\textrm{KL}}(g_W(W | \widetilde{\bmx}_i)\, \| \,f_W(W | \widetilde{\bmx}_i;  \bmtheta, \bmLambda, \bmGamma)}] \leq 1.
\end{equation*}
For a given $\kappa > 0$, define the event
\begin{equation*}
    \mathcal{F}(\kappa) = \qty{\|\bmtheta - \widehat{\bmtheta}_{\textrm{IPD}}\|^2 + \sum_{j=1}^m \{\bmLambda(t_j^*) - \widehat{\bmLambda}_{\textrm{IPD}}(t_j^*)\}^2 + \|\bmGamma - \widehat{\bmGamma}_{\textrm{IPD}}\|^2 \leq n^{-1}\kappa^2}. 
\end{equation*}
By the root-$n$ consistency of the IPD-only approach, for any $\kappa > 0$, there exists a $c_{\kappa} > 0$ such that
\begin{equation*}
     \lim_{n\to\infty} \widehat{\expe}_{\textrm{IPD}} [I\{\mathcal{F}(\kappa)\}]  \geq c_{\kappa}
\end{equation*}
almost surely.
Then, by using Taylor expansion and the law of large numbers, we have
\begin{equation}  \label{eq expe kl2}
    \begin{split}
        &{} \widehat{\expe}_{\textrm{IPD}}\qty[ \exp\qty{ 
        -\sum_{i=1}^n D_{\textrm{KL}}(g_W(W | \widetilde{\bmx}_i)\, \| \,f_W(W | \widetilde{\bmx}_i;  \bmtheta, \bmLambda, \bmGamma)}]       \\ 
        \geq{}& \widehat{\expe}_{\textrm{IPD}}\qty[I\{\mathcal{F}(\kappa)\} \exp\qty{ -\frac{n}{2} A[\bmtheta - \bmtheta_0, \bmLambda- \bmLambda_0,\bmGamma - \bmGamma_0]^{\otimes 2} }] \times (1+o(1)) \\
        \geq{}& \pr(\mathcal{F}(\kappa)) \inf_{(\bmtheta, \bmLambda, \bmGamma) \in \mathcal{F}(\kappa)} \qty[\exp\qty{ -\frac{n}{2} A[\bmtheta - \bmtheta_0, \bmLambda- \bmLambda_0,\bmGamma - \bmGamma_0]^{\otimes 2} }]
        \times (1+o(1))
    \end{split}
\end{equation}
almost surely.
For the IPD-only estimate, we have
\begin{equation*}
    n \qty[ \|\widehat{\bmtheta}_{\textrm{IPD}} - \bmtheta_0\|^2 + \sum_{j=1}^m \{\widehat{\bmLambda}_{\textrm{IPD}}(t_j^*) - \bmLambda_0(t_j^*)\}^2 + \|\widehat{\bmGamma}_{\textrm{IPD}} - \bmGamma_0\|^2 ]= O(1)
\end{equation*}
almost surely.
So we further have
\begin{equation*}
     \lim_{n\to \infty} \inf_{(\bmtheta, \bmLambda, \bmGamma) \in \mathcal{F}(\kappa)} \qty[\exp\qty{ -\frac{n}{2} A[\bmtheta - \bmtheta_0, \bmLambda- \bmLambda_0,\bmGamma - \bmGamma_0]^{\otimes 2} }] > 0.
\end{equation*}
almost surely.
By substituting this into \eqref{eq expe kl2}, we can conclude that the event $\mathcal{E}_n$ holds almost surely as $n \to \infty$.

Since $\mathcal{E}_n$ holds almost surely as $n \to \infty$, it suffices to verify the characteristic function of the posterior distribution converges to that of a Gaussian process when $\mathcal{E}_n$ holds.
Let $\widetilde{\mathcal{H}}$ be the Hilbert space defined on $\mathbb{R}^{p} \times \textrm{BV}_{\tau} \times \mathbb{R}^{p}$ equipped with the norm, for $\widetilde{h} = (h_1, h_2, h_3) \in \widetilde{\mathcal{H}}$,
\begin{equation*}
    \|\widetilde{h}\|_{\widetilde{\mathcal{H}}}
    = \qty[ \|h_1\|^2 + \int_0^\tau [h_2(t)]^2 \dd{t} + \|h_3\|^2]^{1/2}
\end{equation*}
The inner product $\innp{\cdot}{\cdot}_{\widetilde{\mathcal{H}}}$ is defined accordingly.
Let
\begin{equation*}
    \zeta = n^{1/2}(\bmtheta - \widehat{\bmtheta}_{\textrm{IPD}}, \bmLambda- \widehat{\bmLambda}_{\textrm{IPD}}, \bmGamma - \widehat{\bmGamma}_{\textrm{IPD}}),\;
    \zeta_0 = n^{1/2}(\widehat{\bmtheta}_{\textrm{IPD}} - \bmtheta_0, \widehat{\bmLambda}_{\textrm{IPD}} - \bmLambda_0, \widehat{\bmGamma}_{\textrm{IPD}} - \bmGamma_0).
\end{equation*}
Then by using Taylor expansion up to the second order and convexity of the Kullback--Leibler divergence, we have, when the event $\mathcal{E}_n$ holds,
\begin{equation} \label{eq chf all}
\begin{split}
    & \widehat{\expe} \qty[\exp\{\textrm{i}\innp{\widetilde{h}}{\zeta}_{\widetilde{\mathcal{H}}}\}  ] \\
    \propto{}& \widehat{\expe}_{\textrm{IPD}} \qty[\exp\qty{\textrm{i}\innp{\widetilde{h}}{ \zeta }_{\widetilde{\mathcal{H}}}
    -\sum_{i=1}^n D_{\textrm{KL}}(g_W(W | \widetilde{\bmx}_i)\, \| \,f_W(W | \widetilde{\bmx}_i;  \bmtheta, \bmLambda, \bmGamma) }]    \\
    \propto{}& \widehat{\expe}_{\textrm{IPD}} \qty[\exp\qty{\textrm{i}\innp{\widetilde{h}}{\zeta }_{\widetilde{\mathcal{H}}}
    -2^{-1}A[ \zeta  + \zeta_0]^{\otimes 2} } ]  \times \{1+o(1)\} ,
\end{split}
\end{equation}

Let $\{\xi_l, \phi_l\}_{l=1}^\infty$ be the series of the pairs of eigenvalue and eigenfunction of $\bmV^{-1}$ such that $\xi_1 \geq \xi_2 \geq \dots \geq 0$ and $\{\phi_l\}_{l=1}^\infty$ is a complete orthogonal basis on $\widetilde{\mathcal{H}}$.
Then there exist two series of expansion coefficients $\{u_l\}_{l=1}^\infty$ and $\{u_{0l}\}_{l=1}^\infty$ such that
\begin{equation*}
    \zeta = \sum_{l=1}^\infty u_l \phi_l = \sum_{l=1}^\infty \innp{\zeta}{\phi_l}_{\widetilde{\mathcal{H}}} \phi_l,\;\;
    \zeta_0 = \sum_{l=1}^\infty u_{0l} \phi_l = \sum_{l=1}^\infty \innp{\zeta_0}{\phi_l}_{\widetilde{\mathcal{H}}} \phi_l .
\end{equation*}
Define the partial sums as $\zeta_r = \sum_{l=1}^r u_l \phi_l $, $\zeta_{0r} = \sum_{l=1}^r u_{0l} \phi_l$, $\widetilde{h}_r = \sum_{l=1}^r \innp{\widetilde{h}}{\phi_l}_{\widetilde{\mathcal{H}}}  \phi_l$ and $\bmV_r^{-1}(s,t) = \sum_{l=1}^r \zeta_l \phi_l(s) \phi_l(t)$ for $s,t \in [0,\tau]$.
Then we have $\|\zeta_r - \zeta\|_{\widetilde{\mathcal{H}}} \downarrow 0$ and $\bmA[ \zeta_r  + \zeta_{0r}]^{\otimes 2} \uparrow \bmA[ \zeta  + \zeta_0]^{\otimes 2}$ monotonically as $r \to \infty$.
Note that $|\exp\{\textrm{i}\innp{\widetilde{h}}{\zeta }_{\widetilde{\mathcal{H}}}-2^{-1}\bmA[ \zeta  + \zeta_0]^{\otimes 2} \}| \leq 1$.
By using the bounded convergence theorem, we have
\begin{equation*}
    \widehat{\expe}_{\textrm{IPD}} \qty[\exp\qty{\textrm{i}\innp{\widetilde{h}}{\zeta }_{\widetilde{\mathcal{H}}}
    -2^{-1}\bmA[ \zeta  + \zeta_0]^{\otimes 2} } ]
    = \lim_{r\to\infty} \widehat{\expe}_{\textrm{IPD}} \qty[\exp\qty{\textrm{i}\innp{\widetilde{h}_r}{\zeta_r }_{\widetilde{\mathcal{H}}}
    -2^{-1}\bmA[ \zeta_r  + \zeta_{0r}]^{\otimes 2} } ].
\end{equation*}
Direct calculation yields
\begin{equation*}
     \widehat{\expe}_{\textrm{IPD}} \qty[\exp\qty{\textrm{i}\innp{\widetilde{h}_r}{\zeta_r }_{\widetilde{\mathcal{H}}}
    -2^{-1}\bmA[ \zeta_r  + \zeta_{0r}]^{\otimes 2} } ]
    \propto
    \exp[-\textrm{i}\innp{\widetilde{h}_r}{(\bmA+\bmV_r^{-1})^{-1}\bmA\zeta_{0r}}_{\widetilde{\mathcal{H}}} - 2^{-1} \innp{\widetilde{h}_r}{(\bmA+\bmV_r^{-1})\widetilde{h}_r}_{\widetilde{\mathcal{H}}}].
\end{equation*}
By letting $r \to \infty$, we can conclude that
\begin{equation*}
     \widehat{\expe}_{\textrm{IPD}} \qty[\exp\qty{\textrm{i}\innp{\widetilde{h}}{\zeta }_{\widetilde{\mathcal{H}}} -
    2^{-1}\bmA[ \zeta  + \zeta_0]^{\otimes 2} } ]
    \propto
    \exp[-\textrm{i}\innp{\widetilde{h}}{(\bmA+\bmV^{-1})^{-1}\bmA\zeta_{0}}_{\widetilde{\mathcal{H}}} - 2^{-1} \innp{\widetilde{h}}{(\bmA+\bmV^{-1})\widetilde{h}}_{\widetilde{\mathcal{H}}}]
\end{equation*}
Substituting this into \eqref{eq chf all} yields
\begin{equation*}
    \widehat{\expe} \qty[\exp\{\textrm{i}\innp{\widetilde{h}}{\zeta}_{\widetilde{\mathcal{H}}}\} ] \propto
    \exp[-\textrm{i}\innp{\widetilde{h}}{(\bmA+\bmV^{-1})^{-1}\bmA\zeta_{0}}_{\widetilde{\mathcal{H}}} - 2^{-1} \innp{\widetilde{h}}{(\bmA+\bmV^{-1})\widetilde{h}}_{\widetilde{\mathcal{H}}}] \times \{1+o(1)\}.
\end{equation*}
Then it immediately follows that
\begin{equation*}
    \zeta \to \textrm{GP} (-(\bmA+\bmV^{-1})^{-1}\bmA\zeta_{0},\, (\bmA+\bmV^{-1})^{-1})
\end{equation*}
weakly as $n \to \infty$.
By combining this with \eqref{eq mle}, we further have
\begin{equation*}
    n^{1/2}(\bmtheta - \widehat{\bmtheta}, \bmLambda- \widehat{\bmLambda}, \bmGamma - \widehat{\bmGamma}) =  \zeta + \zeta_0 \to  \textrm{GP} (\bmzero,(\bmA+\bmV^{-1})^{-1})
\end{equation*}
weakly as $n \to \infty$.

Step 3. By an argument similar to the proof of equation (10) of \cite{Kim2006Bernsteinvon}, we can prove that $\widehat{\expe}_{\textrm{IPD}}\{n^{1/2}(\bmtheta - \widehat{\bmtheta}_{\textrm{PL}})\} < \infty$ holds in probability for large enough $n$.
By using the expression of the IPD-only posterior distribution for $\bmLambda$ in Theorem 2.1 of \cite{Kim2006Bernsteinvon}, it is not hard to verify that $\widehat{\expe}_{\textrm{IPD}}\{n^{1/2}(\bmLambda - \widehat{\bmLambda}_{\textrm{PL}})\} < \infty$ for all $n$.
Since the event $\mathcal{E}_n$ defined in \eqref{ev e} holds almost surely, we can further conclude that $\widehat{\expe}[n^{1/2}(\bmtheta - \widehat{\bmtheta}, \bmLambda- \widehat{\bmLambda}, \bmGamma - \widehat{\bmGamma})] < \infty$ holds in probability for large enough $n$, which completes the proof.

\subsection{Proof of Corollary 1}
Recall that the covariance function of the IPD-only estimate of $\bmtheta$ is $\bmV$.
By Theorem 1, the asymptotic covariance function of the maximum likelihood estimator and the posterior variance function of the proposed approach are, respectively, given by
\begin{equation*}
    \bmSigma_1 = (\bmV^{-1}+\bmA)^{-1} \bmV^{-1} (\bmV^{-1}+\bmA)^{-1},\;\;
    \bmSigma_2 = (\bmV^{-1}+\bmA)^{-1}. 
\end{equation*}
Then, it immediately follows that $\bmSigma_1 =  \bmSigma_2 \bmV^{-1} \bmSigma_2$.

\subsection{Theorem 2 and Proof}

\subsubsection{Theoretical results (Theorem 2) for baseline hazards}

We have restricted the increments of the baseline hazard to occur only at the set of observed event times, $\mathcal{T}_0 = \{y_i: \Delta_i = 1,\, i=1,\dots,\, n\}$. Under the internal-only proportional hazards model, the full likelihood is maximized when the baseline hazard has increments exclusively at $\mathcal{T}_0 $. In the proposed framework, which incorporates predictions at time points $\{t_w^*\}_{w=1}^m$ that may not coincide with elements of $\mathcal{T}_0$, it is natural to ask whether this property still holds. The following theorem addresses this question.


\setcounter{theorem}{1}
\begin{theorem} \label{theo loc jumps 2}
(a) There exists a stepwise constant baseline cumulative hazard function with increments located at most at points in $\mathcal{T}_0  \cup \{t_w^*\}_{w=1}^m$ that maximizes $\{\pi(\bmtheta, \bmLambda, \bmGamma) \}^{-1} \pi(\bmtheta, \bmLambda, \bmGamma | \mathcal{D}_n, g_W) $.
(b) Suppose that the conditions of Theorem 1 in the main manuscript hold.
    In addition, suppose that the cumulative distribution function of $C$ is continuous in the neighborhood of $ \{t_w^*\}_{w=1}^m$. 
    Then, the baseline hazard function maximizing  $\{\pi(\bmtheta, \bmLambda, \bmGamma) \}^{-1} \pi(\bmtheta, \bmLambda, \bmGamma | \mathcal{D}_n, g_W) $ has increments only on $\mathcal{T}_0$ in probability as $n\to\infty$.
\end{theorem}

Part~(a) provides a finite-sample characterization of the support of the maximizing baseline hazard,  while part~(b) shows that, asymptotically and under mild additional conditions, the increments are confined to $\mathcal{T}_0$. The additional condition in part (b) of Theorem \ref{theo loc jumps 2} is satisfied if the distribution of C is continuous. Therefore, by Theorem~2, it is sufficient in most cases to restrict the set of increment locations to $\mathcal{T}_0$ for computation. 

\subsubsection{Proof of Theorem 2}

We first consider Theorem 2(a).
Suppose there exists another stepwise constant baseline cumulative hazard function $\bmLambda^\star$ that maximizes $\ell(\bmtheta, \bmLambda, \bmGamma)$, the log-likelihood function of the proposed approach, and has an increment of size $\lambda^\star$ at some time point $t^\star \notin \mathcal{T}$.
  	First, if $t^\star$ is larger than all time points in $\mathcal{T}_0  \cup \{t_w^*\}_{w=1}^m$, the increment at $t^\star$ can be removed without decreasing  $\ell(\bmtheta, \bmLambda, \bmGamma)$.
  	Suppose that $t^\star$ is smaller than some points in $\mathcal{T}_0  \cup \{t_w^*\}_{w=1}^m$.
  	Then, removing the increment at $t^\star$ and adding $\lambda^\star$ to the increment at $\argmin \{t \in \mathcal{T}_0  \cup \{t_w^*\}_{w=1}^m : t > t^\star\}$ (which may possibly be zero) still does not decrease the external prediction-informed likelihood.
  	This completes the proof of Theorem 2(a).

We now consider Theorem 2(b).
Define the events
\begin{equation*}
    \mathcal{F}_1 = \{\textrm{No ties among $\{Y_i\}_{i=1}^n$}\},\; 
    \mathcal{F}_2 = \{\textrm{No fully observed event times at $\{t_j^*\}_{j=1}^m$}\}.
\end{equation*}
By Condition (C3), it follows that $\mathcal{F}_1 \cap \mathcal{F}_2$ holds almost surely.
Therefore, without loss of generality, we assume that $\mathcal{F}_1 \cap \mathcal{F}_2$ always holds in the remainder of the proof.

We prove the theorem by contradiction.
Let $(\widehat{\bmtheta}, \widehat{\bmLambda}, \widehat{\bmGamma})$ denote the optimal point that maximizes the objective function $\ell$.
Suppose that $\widehat{\bmLambda}$ has at least one positive increment at some point $t_j^*$.
Here, the idea is to prove that there exists a direction of baseline hazard function $\widetilde{\bmLambda}$ with $\dd\widetilde{\bmLambda}(t_j^*) < 0$ such that $\widehat{\bmLambda} + \varepsilon\widetilde{\bmLambda}$ remains in the feasible region for small $\varepsilon > 0$ and
\begin{equation} \label{eq grad}
    \lim_{\varepsilon \downarrow 0} 
    \frac{1}{\varepsilon}\{\ell(\widehat{\bmtheta}, \widehat{\bmLambda} + \varepsilon\widetilde{\bmLambda}, \widehat{\bmGamma}) - \ell(\widehat{\bmtheta}, \widehat{\bmLambda}, \widehat{\bmGamma})\} > 0.
\end{equation}
Then if \eqref{eq grad} holds, the objective function $\ell$ cannot be maximized at $\widehat{\bmLambda}$, which is a contradiction.
As a consequence, we can conclude the optimal $\widehat{\bmLambda}$ must lie on the boundary of the feasible region, which implies that the increment sizes at $\{t_j^*\}_{j=1}^m$ are zero.

For simplicity, we only prove that there is no increment at $t_1^*$.
The theorem follows by repeating the proof for the other $t_j^*$.
Suppose that the increment at $t_1^*$, denoted by $\widehat{\lambda^*}$, is nonzero.
Let the $\widehat{\lambda}_{k_1}$ denote the size of the increment located at $t_{k_1}$, the largest complete observed survival time less than $t_1^*$.
We can construct the direction function $\widetilde{\bmLambda}$ as follows:
\begin{equation*}
    \widetilde{\bmLambda}_{k_1} = 1-\widehat{\lambda}_{k_1};\;
    \widetilde{\bmLambda}^* = -(1-\widehat{\lambda}^*);\;
    \widetilde{\bmLambda}(t) =  0,\; \textrm{ for }
    t \in [0, t_{k_1}) \cup [t_1^*, \tau].
\end{equation*}
By this construction, it is easy to verify  
\begin{equation*}
     \lim_{\varepsilon \downarrow 0} [ \varepsilon^{-1} D_{\textrm{KL}}\{g_W(W | \widetilde{\bmX})\, \| \,f_W(W | \widetilde{\bmX};  \bmtheta, \widehat{\bmLambda} + \varepsilon\widetilde{\bmLambda}, \bmGamma)\} ] = 0
\end{equation*}
for any $\bmtheta$ and $\bmGamma$.
Therefore, in order to prove \eqref{eq grad}, it suffices to prove
\begin{equation} \label{eq grad 2}
     \lim_{\varepsilon \downarrow 0}  \frac{\ell_{\textrm{PH}}(\widehat{\bmtheta}, \widehat{\bmLambda} + \varepsilon\widetilde{\bmLambda}) - \ell_{\textrm{PH}}(\widehat{\bmtheta}, \widehat{\bmLambda} )}{\varepsilon} > 0.
\end{equation}
When $\mathcal{F}_1 \cap \mathcal{F}_2$ holds, direct calculation yields
\begin{equation*}
    \pdv{\ell_{\textrm{PH}}(\widehat{\bmtheta}, \widehat{\bmLambda}) }{\widehat{\lambda}_{k_1}}
    = -\sum_{i: Y_i > t_{k_1}} e^{\widehat{\bmtheta}^T \widetilde{\bmx}_i} \frac{1}{1 - \widehat{\lambda}_{k_1}}
        + \frac{e^{\widehat{\bmtheta}^T \widetilde{\bmx}_{k_1}} (1 - \widehat{\lambda}_{k_1})^{e^{\widehat{\bmtheta}^T \widetilde{\bmx}_{k_1}} - 1 }}{1 -  (1 - \widehat{\lambda}_{k_1})^{e^{\widehat{\bmtheta}^T \widetilde{\bmx}_{k_1}} }}, 
\end{equation*}
\begin{equation*}
    \pdv{\ell_{\textrm{PH}}(\widehat{\bmtheta}, \widehat{\bmLambda}) }{\widehat{\lambda}^*}
    = -\sum_{i: Y_i > t_1^*} e^{\widehat{\bmtheta}^T \widetilde{\bmx}_i} \frac{1}{1 -\widehat{\lambda}^*} .
\end{equation*}
Then we further have
\begin{equation} \label{eq grad 3}
    \begin{split}
    & \lim_{\varepsilon \downarrow 0}  \frac{\ell_{\textrm{PH}}(\widehat{\bmtheta}, \widehat{\bmLambda} + \varepsilon\widetilde{\bmLambda}) - \ell_{\textrm{PH}}(\widehat{\bmtheta}, \widehat{\bmLambda}) }{\varepsilon}\\
    ={}&  ( 1-\widehat{\lambda}_{k_1})  \pdv{\ell_{\textrm{PH}} (\widehat{\bmtheta}, \widehat{\bmLambda}) }{\widehat{\lambda}_{k_1}} - (1-\widehat{\lambda}^*) \sum_{i: Y_i > t_1^*} e^{\widehat{\bmtheta}^T \widetilde{\bmx}_i} \frac{1}{1 -\widehat{\lambda}^*}  \\
    ={}& -\sum_{i: t_{k_1} < Y_i \leq t_1^*} e^{\widehat{\bmtheta}^T \widetilde{\bmx}_i} +  \frac{e^{\widehat{\bmtheta}^T \widetilde{\bmx}_{k_1}} (1 - \widehat{\lambda}_{k_1})^{e^{\widehat{\bmtheta}^T \widetilde{\bmx}_{k_1}} }}{1 -  (1 - \widehat{\lambda}_{k_1})^{e^{\widehat{\bmtheta}^T \widetilde{\bmx}_{k_1}} }} 
    = -S_1 + S_2.
    \end{split}
\end{equation}
Now, in order to prove \eqref{eq grad 2}, our aim is to provide the magnitude of $S_1$ and $S_2$ on the right-hand side of \eqref{eq grad 3}.

We first consider $S_1$.
By Condition (C3), there exists a $\kappa > 0$ such that
\begin{equation*}
    \pr(t_1^* - t_{k_1} \geq n^{-2/3}) = \prod_{i=1}^n \{ 1 - \pr(T_i \in (t_1^* - n^{-2/3}, t_1^*],\; \Delta_i = 1) \} \leq (1 - \kappa n^{-2/3})^n .
\end{equation*}
Consequently, the event $\mathcal{F}_3 = \{t_1^* - t_{k_1} < n^{-2/3}\}$ holds in probability.
Then, by the additional condition of this theorem, we can conclude that 
\begin{equation*}
    \sum_{i=1}^n I[ t_{k_1} < Y_i \leq t_1^* ] I[\Delta_i = 1] = O_p(n^{1/3}), 
\end{equation*}
which yields $S_1  = O_p(n^{1/3})$.

We now consider $S_2$.
By Theorem 1, we have
\begin{equation} \label{eq L tstar}
    \widehat{\bmLambda}(t_1^*) - \bmLambda(t_1^*)  = O_p(n^{-1/2})
\end{equation} 
On the other hand, the cumulative baseline hazard function is bounded, continuous, and monotonically increasing on $[0,\tau]$ by Conditions (C2) and (C3).
By using the continuous mapping theorem, we can conclude that 
\begin{equation*}
     \widehat{\bmLambda}\{(t_1^*)^-\} - \bmLambda(t_1^*) = O_p(n^{-1/2})
\end{equation*}
Combining this with \eqref{eq L tstar} yields 
\begin{equation*}
    \widehat{\lambda^*} = \widehat{\bmLambda}(t_1^*) - \widehat{\bmLambda}\{(t_1^*)^-\} = O_p(n^{-1/2}).
\end{equation*}
So we have
\begin{equation*}
    1 -  (1 - \widehat{\lambda}_{k_1})^{e^{\widehat{\bmtheta}^T \widetilde{\bmx}_{k_1}} } = O_p(n^{-1/2}),
\end{equation*}
and hence $S_1^{-1} S_2 \to \infty$ in probability, which completes the proof of \eqref{eq grad 2}.

\subsection{Variance Correction}

For the $l$th entry of $\bmtheta$, denoted by $\bmtheta_l$, its asymptotic variance is given by
\begin{equation*}
   \bmSigma_{1,ll} = \bme_l^\top \bmSigma_1 \bme_l
    = V^{-1}[\bmSigma_2 \bme_l]^{\otimes 2}, 
\end{equation*}
where $\bme_l$ is the unit vector associated with $\bmtheta_l$.
Let $\bmlambda = (\lambda_1,\dots,\,\lambda_J)$.
Let $\bmU$ denote the operator mapping  
$(\bmtheta, \bmlambda, \bmGamma) \mapsto (\bmtheta, \bmLambda,\bmGamma)$
Then the estimate of $\bmSigma_{1,ll}$ can be expressed as
\begin{equation*}
    \widehat{\bmSigma}_{1,ll} =  (\bmU\widehat{\bmV}\bmU)^{-1}[(\bmU\widehat{\bmSigma}_2 \bmU) (\bmU^{-1}\bme_l)]^{\otimes 2}
                = \widetilde{\bmV}^{-1}[\widetilde{\bmSigma}_2 \widetilde{\bme}_l]^{\otimes 2} .
\end{equation*}
where $\widetilde{\bmV}^{-1}$ is the sample information operator with respect to $ (\bmtheta, \bmlambda, \bmGamma) $ for the IPD-only approach, and $\widetilde{\bmSigma}_2$ can be estimated by the posterior covariance of $(\bmtheta, \bmlambda, \bmGamma)$.
We divide the variance correction computation into two steps.

Step 1. We first consider $\widetilde{\bmSigma}_2 \widetilde{\bme}_l$.
Note that $\widetilde{\bme}_l = \bmU^{-1}\bme_l = \bme_l$.
Let $\widetilde{\bmSigma}_2 \widetilde{\bme}_l = (h_{\theta}, h_{\lambda}, h_{\Gamma})$.
Then $h_{\theta}$, $h_{\lambda}$ and $h_{\Gamma}$ can be respectively estimated by the posterior covariance $\bmOmega_{\theta,l} = \widehat{\cov}(\bmtheta, \bmtheta_l)$, 
$\bmOmega_{\Lambda,l} = \widehat{\cov}(\bmlambda, \bmtheta_l)$ and  $\bmOmega_{\Gamma,l} = \widehat{\cov}(\bmGamma, \bmtheta_l)$.

Step 2. We next consider $\widetilde{\bmV}^{-1}$.
Let $(\overline{\bmlambda}, \overline{\bmtheta})$ denote the posterior mean of the proposed approach. 
Recall that $\widetilde{\bmV}^{-1}$ is block-diagonal with the cross term corresponding to $(\bmtheta, \bmlambda)$ and $\bmGamma$ is zero.
So we have
\begin{equation*}
	\widehat{\bmV}_{ll}^{-1} = \widehat{\sigma}_{\theta\theta}[h_{\theta}]^{\otimes 2} + 2\widehat{\sigma}_{\theta\lambda}[h_{\theta}][h_{\Lambda}] + \widehat{\sigma}_{\Lambda\Lambda}[h_{\Lambda}]^{\otimes 2} + \widehat{\bmSigma}_{\Gamma\Gamma}^{-1} [h_{\Gamma}]^{\otimes 2}.
\end{equation*} 
Here $\widehat{\bmSigma}_{\Gamma\Gamma}^{-1}$ can be estimated using the standard result for the submodel of $Z \mid \bm{X}$.

We now consider $\widehat{\sigma}_{\theta\theta}$, $\widehat{\sigma}_{\theta\lambda}$ and $\widehat{\sigma}_{\lambda\lambda}$.
In Section \ref{sec MLE}, we have proved $\|\widehat{\bmtheta}_{\textrm{IPD}} - \widehat{\bmtheta}_{\textrm{PL}}\| = O_p(n^{-1})$ and 
$|\dd\widehat{\bmLambda}_{\textrm{IPD}}(t) - \dd\widehat{\bmLambda}_{\textrm{PL}}(t)| = o_p(n^{-3/2})$.
Therefore, we can use the information operator of $(\widehat{\bmtheta}_{\textrm{PL}}, \widehat{\bmlambda}_{\textrm{PL}})$
as the estimate of the information operator of $(\widehat{\bmtheta}_{\textrm{IPD}} , \widehat{\bmlambda})$.
It is known that the Cox partial likelihood is the profile likelihood of the full likelihood with respect to the baseline hazard. 
See \cite{Murphy2000profile} for more details.
Therefore, the information operator of $(\widehat{\bmtheta}_{\textrm{PL}}, \widehat{\bmlambda}_{\textrm{PL}})$ can be derived from the full likelihood function of the proportional hazards model, whose logarithm is given by
\begin{equation*}
    \ell(\bmtheta, \bmLambda)
= \sum_{i=1}^n \int_0^\tau
\left[ \bmtheta^\top \widetilde{\bmx}_i  \dd{N_i(t)} + \log \dd{\bmLambda(t)} \dd{N_i(t)}
- R_i(t)\exp(\bmtheta^\top \widetilde{\bmx}_i) \dd{\bmLambda}(t)
\right],
\end{equation*}
where $N_i(t) = I[Y_i \leq t, \Delta_i = 1]$ is the counting process, and $R_i(t) = I[Y_i \geq t]$ is the at-risk process.
By taking the partial derivative, the score functions are respectively given by
\begin{equation*}
    \pdv{\ell(\bmtheta, \bmLambda)}{\bmtheta} = \sum_{i=1}^n \int_0^\tau
\left[ \widetilde{\bmx}_i \, \dd{N_i(t)} 
- \widetilde{\bmx}_i R_i(t)\exp(\bmtheta^\top \widetilde{\bmx}_i)\, \dd{\bmLambda}(t)
\right] =  \sum_{i=1}^n \int_0^\tau \widetilde{\bmx}_i  \dd{M_i(t)} ,
\end{equation*}
\begin{equation*}
    \pdv{\ell(\bmtheta, \bmLambda)}{({\dd{\bmLambda}(t)})} = \sum_{i=1}^n 
\left[ \frac{\dd{N_i(t)}}{ \dd{\bmLambda}(t)}
- R_i(t)\exp(\bmtheta^\top \widetilde{\bmx}_i)
\right] = \sum_{i=1}^n \int_0^\tau \frac{\dd{M_i(t)}}{\dd{\bmLambda}(t)},
\end{equation*}
where $\dd{M_i(t)} = \dd{N_i(t)} - R_i(t) e^{\bmtheta^\top \widetilde{\bmx}_i} \dd{\bmLambda}(t)$ is the martingale increment, whose predictable variation is given by
\begin{equation*}
    \dd{\langle M_i \rangle}(t) = Y_i(t) e^{\bmtheta^\top \widetilde{\bmx}_i} \dd{\bmLambda}(t) .
\end{equation*}
See, for example, Chapter 2 of \cite{ABGK} for more details about the martingale in survival analysis.
By taking the covariance, we can further have 

\begin{equation*}
	\widehat{\sigma}_{\theta\theta}[h_{\theta}]^{\otimes 2}
	= \frac{1}{n}\sum_{j=1}^m \sum_{i=1}^n \qty{  (h_{\theta}^\top \widetilde{\bmx}_j)^2 R_i(t_j) e^{\overline{\bmtheta}^\top \widetilde{\bmx}_j} } \overline{\lambda}_{t_j} ,
\end{equation*}
\begin{equation*}
	\widehat{\sigma}_{\theta\Lambda}[h_{\theta}][h_{\Lambda}]
	= \frac{1}{n}\sum_{j=1}^m \sum_{i=1}^n \qty{  (h_{\theta}^\top \widetilde{\bmx}_j) R_i(t_j) e^{\overline{\bmtheta}^\top \widetilde{\bmx}_j} } h_{\Lambda}(t_j)  ,
\end{equation*}
\begin{equation*}
	\widehat{\sigma}_{\Lambda\Lambda}[h_{\Lambda}]^{\otimes 2} 
	= \frac{1}{n}\sum_{j=1}^m \sum_{i=1}^n \frac{1}{\overline{\lambda}_{t_j}} R_i(t_j) e^{\overline{\bmtheta}^\top \widetilde{\bmx}_j}  \{h_{\Lambda}(t_j)\}^2 .
\end{equation*}

\vspace{3em} 
\section{Details of model extension and Bayesian computation} 

\subsection{A Working Imputation Model of $\mathbf{Z}$}
We focus on the case where $\bmZ$ is continuous, which covers both our simulation studies and the real-data application.
In this case, we assume $\bmZ|\bmx \sim MVN\{\mathit{E}(\bmZ|\bmx), \sigma^2 \bmI \}$ where $MVN$ denotes the multivariate normal distribution. To induce sparse estimation in the working imputation model, we adopt a Bayesian LASSO specification. Specifically, we assume $ \gamma_{kl} | \zeta_{kl}^2 \sim N(0, \zeta_{kl}^2)$, $\zeta_{kl}^2 | \alpha_k \sim \text{Exp}(\alpha_k/2)$, and $\alpha_k \sim \text{Gamma}(a_\alpha, b_\alpha)$, where Exp denotes the exponential distribution, and Gamma denotes the gamma distribution with shape parameter $a_\alpha$ and rate parameter $b_\alpha$. Marginalizing over $\zeta_{kl}^2$, the induced prior distribution of $\gamma_{kl}$ given $\alpha_k$ follows a Laplace distribution with location parameter zero and scale parameter $1/\sqrt{\alpha_k}$, thereby recovering the Lasso penalty $\sqrt{\alpha_k} | \gamma_{kl}|$ \citep{ParkCasella2008}. Accordingly, $\bmGamma$ denotes the collection of all parameters associated with the working model for $\bmZ$, i.e., $\bmGamma = ( \bmgamma_1, \ldots, \bmgamma_{p_z}, \bmzeta_1,\ldots,\bmzeta_{p_z}, \bmalpha, \sigma^2 )$ where $\bmgamma_k$ and $\bmzeta_k$ are $p_{\gamma_k}-$dimensional vectors of regression coefficients and their variance parameters, respectively, and $\bmalpha=(\alpha_1,\ldots,\alpha_{p_z})^\top$ is the vector of shrinkage hyperparameters.

\subsection{Bayesian Computation}

We describe the Bayesian computation for the proposed framework. 
Posterior inference is conducted using Markov chain Monte Carlo (MCMC) simulation. Incorporating the three extensions described in Section 2.4 from the main text, the joint posterior distribution of all model parameters is given by  
\begin{align}
    & \pi(\bmtheta, \nu, \bmLambda, \bmGamma, g_W | \mathcal{D}_n, \hat{g}_W) \label{eq target posterior} \\
    & \propto \prod_{i=1}^n f( y_i | \Delta_i, \bmx_i, \bmz_i; \bmtheta, \bmLambda ) f(\bmz_i | \bmx_i; \bmGamma) \cdot f(\nu | \mathcal{X}_n; g_W, \bmtheta, \bmLambda, \bmGamma ) \pi(\bmtheta, \nu, \bmLambda, \bmGamma | g_W ) \pi(g_W | \hat{g}_W) \nonumber 
\end{align}
where $\pi(\bmtheta, \nu, \bmLambda, \bmGamma | g_W ) \propto \exp \left\{ - \sum_{i=1}^{n}  D_{KL} \left( g_{W}(\cdot|\bmx_i) \| f_{W}( \cdot | \bmx_i ; \bmtheta, \nu, \bmLambda, \bmGamma)  \right) \right\} \pi(\bmtheta, \nu, \bmLambda, \bmGamma)$. 

We assign weakly informative priors to the model parameters. Specifically, we place an independent $N(0, 10^4)$ prior on each component of $\bmtheta$. The cumulative baseline hazard $\bmLambda$ is modeled through a Beta process prior, and we adopt the improper prior $\pi(\lambda_j) \propto 1/\lambda_j$ for the baseline hazard increments $\lambda_j$, $j=1,\ldots,J$ ~\citep{Hjort1990, KimLee2003Bayesian}. For the Bayesian LASSO specification introduced in Web Appendix C.1, we further assume $\sigma^2 \sim \text{Inv-Gamma}(1,1)$ where $\text{Inv-Gamma}$ denotes the inverse-gamma distribution with shape and rate parameters, and $\nu \sim N(0, 10^4)$. The MCMC algorithm is run for 5,000 iterations, with the first 20\% discarded as burn-in.

Given the current parameter values $(\bmtheta, \nu, \bmLambda, \bmgamma_1, \ldots, \bmgamma_{p_z}, \bmzeta_1,\ldots,\bmzeta_{p_z}, \bmalpha, \sigma^2, g_W)$, one MCMC iteration proceeds as follows.

\begin{enumerate}

	\item For each $\theta_j$ for $j=1,\ldots,p_\theta$, generate a candidate value $\theta_j^q$ from a symmetric proposal distribution $q(\theta_j^q|\theta_j)$ and accept it with the probability $\min\{1,Q\}$, where
    \begin{align*}
            Q & = \frac{
		      \prod_{i=1}^{n} f(y_i | \Delta_i, \bmx_i, \bmz_i; \bmtheta^q, \bmLambda) 
	       }{
		      \prod_{i=1}^{n} f(y_i | \Delta_i, \bmx_i, \bmz_i; \bmtheta, \bmLambda) 
	       } 
           \cdot 
           \frac{
		     f(\nu | \mathcal{X}_n; g_W, \bmtheta^q, \bmLambda, \bmGamma ) 
             \pi ( \bmtheta^q, \nu, \bmLambda, \bmGamma | g_W )
           }{
           f(\nu | \mathcal{X}_n; g_W, \bmtheta, \bmLambda, \bmGamma )
		      \pi ( \bmtheta, \nu, \bmLambda, \bmGamma | g_W )
          }
    \end{align*}
	and $\bmtheta^q = (\theta_1,\ldots \theta_{j-1}, \theta_j^q, \theta_{j+1},\ldots,\theta_{p_\theta})$.
    To evaluate the KL divergence term, we approximate it by the Monte Carlo integration  $f_W(w | \bmx_i; \bmtheta, \bmLambda, \bmGamma)$ $ \approx B^{-1} \sum_{b=1}^{B}$ $P(t_{w-1}^{\ast} $ $< Y $ $\leq t_w^\ast $ $| \bmx_i, \bmz^{(b)}, \bmtheta, \bmLambda)$ where $\bmz^{(b)}$ is sampled from $f(\bmz | \bmx_i;\bmGamma)$, where we set $B$=2,000. 
	
		\item For each observed event time $\{y_i : \Delta_i=1\}$, generate a candidate $\lambda_{i}^q$ from a proposal distribution  $q(\lambda_{i}^q|\lambda_{i})$ whose support is (0,1), then accept it with probability $\min\{1,Q\}$, where
        \begin{equation*}
        \begin{split}
             Q & = \frac{
		      {\prod_{i=1}^{n} f(y_i | \Delta_i, \bmx_i, \bmz_i; \bmtheta, \bmLambda^q) } 
	       }{
		      {\prod_{i=1}^{n} f(y_i | \Delta_i, \bmx_i, \bmz_i; \bmtheta, \bmLambda) } 
	       } \cdot
           \frac{
		      f(\nu | \mathcal{X}_n; g_W, \bmtheta, \bmLambda^q, \bmGamma ) \pi ( \bmtheta, \nu, \bmLambda^q, \bmGamma | g_W ) 
           }{
		      f(\nu | \mathcal{X}_n; g_W, \bmtheta, \bmLambda, \bmGamma ) \pi ( \bmtheta, \nu, \bmLambda, \bmGamma | g_W ) 
           } 
           \cdot
           \frac{
            q(\lambda_{i}|\lambda_{i}^q)
           }{
            q(\lambda_{i}^q|\lambda_{i})
           }
        \end{split}. 
        \end{equation*}
        Here, $\bmLambda^q = (\lambda_1,\ldots,\lambda_{i-1},\lambda_{i}^q,\lambda_{i+1},\ldots,\lambda_{n})$ ordered according to event time. 

        The likelihood ratio (the first term) simplifies to 
			$$ \frac{
				[1 - (1-\lambda_i^q)^{\exp(\bmtheta^T \tilde\bmx_i)} ] \prod_{ \{ i': i' > i \} } (1-\lambda_{i}^{q})^{\exp(\bmtheta^T \tilde\bmx_{i'})} 
			}{
				[ 1 - (1-\lambda_i)^{\exp(\bmtheta^T \tilde\bmx_i)} ] \prod_{ \{ i': i' > i \} } (1-\lambda_{i})^{\exp(\bmtheta^T \tilde\bmx_{i'})} }, $$ 
        where $\tilde\bmx_i^T = (\bmx_i^T, \bmz_i^T)$. 

	\item Generate a candidate $\nu^q$ from a symmetric proposal distribution and accept it with probability $\min\{1,Q\}$, where
	   \begin{equation*}
        \begin{split}
            Q & = \frac{
		      f(\nu^q | \mathcal{X}_n; g_W, \bmtheta, \bmLambda, \bmGamma ) 
              \pi ( \bmtheta, \nu, \bmLambda, \bmGamma^q | g_W )
	       }{
		      f(\nu | \mathcal{X}_n; g_W, \bmtheta, \bmLambda, \bmGamma )  
              \pi ( \bmtheta, \nu, \bmLambda, \bmGamma | g_W )
	       } 
        \end{split}.
        \end{equation*}

	\item Generate a candidate external prediction distribution $g_W^q$ from its proposal distribution $q(g_W^q|g_W)$ and accept it with  probability $\min\{1,Q\}$, where
    \begin{equation*}
        \begin{split}
            Q & = \frac{
		      f(\nu | \mathcal{X}_n; g_W^q, \bmtheta, \bmLambda, \bmGamma ) \pi(\bmtheta, \nu, \bmLambda, \bmGamma | g_W^q ) \pi(g_W^q | \hat{g}_W)
	       }{
		      f(\nu | \mathcal{X}_n; g_W, \bmtheta, \bmLambda, \bmGamma ) \pi(\bmtheta, \nu, \bmLambda, \bmGamma | g_W ) \pi(g_W | \hat{g}_W)
	       } \cdot \frac{
             q ( g_W | g_W^q )
           }{ 
            q ( g_W^q | g_W )
           }
        \end{split}. 
    \end{equation*}
	
	\item For each $\gamma_{kl}$, $k=1,\ldots,p_{z}$ and $l=0,1,\ldots,p_{\gamma_k}$, generate a candidate $\gamma_{kl}^q$ from its symmetric proposal distribution $q(\gamma_{kl}^q|\gamma_{kl})$ and accept it with probability $\text{min}\{1,Q\}$, where 
    \begin{equation*}
        \begin{split}
            Q & = \frac{
		      \prod_{i=1}^{n} f( \bmz_i | \bmx_i, \bmGamma^q)  
	       }{
		      \prod_{i=1}^{n} f( \bmz_i | \bmx_i, \bmGamma) 
	       }  \cdot \frac{
		      f(\nu | \mathcal{X}_n; g_W, \bmtheta, \bmLambda, \bmGamma^q ) \pi ( \bmtheta, \nu, \bmLambda, \bmGamma^q | g_W ) \pi(\gamma_{kl}^q | \zeta_{kl}^2)
           }{
		      f(\nu | \mathcal{X}_n; g_W, \bmtheta, \bmLambda, \bmGamma ) 
              \pi ( \bmtheta, \nu, \bmLambda, \bmGamma | g_W ) \pi(\gamma_{kl} | \zeta_{kl}^2)
           }
        \end{split}
    \end{equation*}
	and $\bmGamma_j^q = \{ (\gamma_{10}, \gamma_{11}, \ldots, \gamma_{kl-1}, \gamma_{kl}^q, \gamma_{kl+1}, \ldots, \gamma_{p_z p_{\gamma_k}}), \bmzeta_1,\ldots,\bmzeta_{p_z}, \bmalpha, \sigma^2 \} $.

    \item Update $\zeta_{kl}^2$, $k=1,\ldots,p_z$ and $l=0,\ldots,p_{\gamma_k}$, using Gibbs sampling from the full-conditional distribution 
    \begin{equation*}
        \zeta_{kl}^2 | \bmgamma_k, \alpha_k \sim \text{Inverse-Gauss} \left(\sqrt{ \frac{\alpha_k}{\sum_{l=0}^{p_{\gamma_k}} \gamma_{kl}^2} }, \alpha_k \right), \,
    \end{equation*}
    where Inverse-Gauss$(a,b)$ denotes the inverse Gaussian distribution with mean $a$ and shape parameter $b$. 
    
    \item Update $\alpha_k$, $k=1,\ldots,p_z$, using Gibbs sampling from its full-conditional distribution 
    \begin{equation*}
        \alpha_k | \zeta_{kl}^2 \sim \text{Gamma} \left( a_\alpha + p_{\gamma_k}, b_\alpha + \frac{1}{2}\sum_{l=0}^{p_{\gamma_k}} \zeta_{kl}^2 \right) \,.
    \end{equation*}
    
	\item Generate a candidate $\sigma^{2q}$ from a symmetric proposal distribution $q(\sigma^{2q}|\sigma^2)$. If $\sigma^{2q}>0$, accept it with probability $\text{min}\{1,Q\}$, where 
    \begin{equation*}
        \begin{split}
            Q & = \frac{
		      \prod_{i=1}^{n} f( \bmz_i | \bmx_i, \bmGamma^q) 
	       }{
		      \prod_{i=1}^{n} f( \bmz_i | \bmx_i, \bmGamma) 
	       }  \cdot \frac{
           f(\nu | \mathcal{X}_n; g_W, \bmtheta, \bmLambda, \bmGamma^q ) 
		      \pi ( \bmtheta, \nu, \bmLambda, \bmGamma^q | g_W ) \pi(\sigma^{2q})
           }{
           f(\nu | \mathcal{X}_n; g_W, \bmtheta, \bmLambda, \bmGamma ) 
		      \pi ( \bmtheta, \nu, \bmLambda, \bmGamma | g_W ) \pi(\sigma^2)
           }
        \end{split}
    \end{equation*}
    where $\bmGamma^q = (\bmgamma_1,\ldots,\bmgamma_{p_z}, \bmzeta_1,\ldots,\bmzeta_{p_z}, \bmalpha, \sigma^{2q} )$.

\end{enumerate}

\vspace{3em} 
\section{Simulation}

\subsection{Variability of External Information}
 
This section describes how the external prediction model and its sampling variability were constructed in the simulation study of main text (Scenario 2).

Based on the data-generating process described in Section 3.2 of the main text, we generated an external dataset of size $n_{ext} = 5,000$. For each observation $j=1,\ldots,n_{ext}$, we defined the discrete survival interval indicator $w_j$ such that $w_j = w$ if $t_{w-1}^\ast < t_j \leq t_{w}^\ast$. Let $\mathcal{D}_{ext}=\{(w_{j},\bmx_{j})\}_{j=1}^{n_{ext}}$ denote the external dataset. Using $\mathcal{D}_{ext}$, we first fitted a multinomial logistic regression model with spline terms, denoted by $\mathcal{G}$, to estimate the probabilities of three disjoint time intervals $I_w=(t_{w-1}^*,t_w^*]$ for $w=1,2,3$. We then applied $\mathcal{G}$ to each covariate vector $\bmx_i$ in the internal dataset for $i=1,\ldots,n$, obtaining the predicted interval probabilities $\hat{g}_W(w,\bmx_i)$.

Next, we calculated the sampling variability of the external predictions using the bootstrap method. Specifically, for each bootstrap iteration $b=1,\ldots,30$, we generated a bootstrap sample $\mathcal{D}_{ext}^{(b)}$ of size $n_{ext}$ by sampling with replacement from $\mathcal{D}_{ext}$. We then refitted the same multinomial logistic regression model to $\mathcal{D}_{ext}^{(b)}$, yielding the bootstrap model $\mathcal{G}^{(b)}$. Using $\mathcal{G}^{(b)}$, we computed the corresponding predicted probabilities $\hat{g}_W^b(w,\bmx_i)$ for the same set of internal covariates $\bmx_i$. Based on the bootstrap replicates, we estimated the covariance matrix of the external prediction information, denoted by $\bmSigma_{\hat{g}}(w,\bmx_i)$. The diagonal elements of $\bmSigma_{\hat{g}}(w,\bmx_i)$  were used as variance estimates for the external prediction information. Finally, following Section 2.4.3 in the main text, we specified the prior distribution of $g_W(w,\bmx_i)$ using the estimated $\hat{g}_W(w,\bmx_i)$ and variance estimate $\widehat{\bmV}(\hat{g}(w,\bmx_i))$ for each $\bmx_i$.

\subsection{Simulation Tables}

\begin{table}[H]
\centering
\caption{Summary of estimated parameters in the working imputation model under Scenario 1 that assumes external prediction with certainty and homogeneous data distribution. We report bias, mean squared error (MSE), and coverage probability (CP) for the external prediction informed Cox (EPI) model.} \label{tab:simul 1_Gamma}

\vspace{1em}
\begin{tabular}{l rrr }
\hline
             & bias & MSE & CP  \\
\hline 
$\gamma_0$   &  0.002 & 0.002 & 0.953  \\
$\gamma_1$   & -0.002 & 0.013 & 0.957  \\
$\gamma_2$   & -0.003 & 0.012 & 0.960  \\
$\sigma^2$   &  0.019 & 0.001 & 0.913  \\
\hline 
\end{tabular}
\end{table}

\begin{table}[H]
\centering
\caption{Summary of estimated parameters in the working imputation models under Scenario 2 that assumes external prediction with sampling errors and heterogeneous data distribution. We report bias, mean squared error (MSE), and coverage probability (CP) of the three variants of the proposed approach; EPI, the fully extended version of the proposed external prediction informed Cox model that accommodates heterogeneous baseline hazards, sampling errors in the external predictions, and a working model for Z; EPI-trueZ, which assumes the true imputation model for Z while allowing heterogeneous baseline hazards and sampling errors in the external predictions; and EPI-oracle, an oracle version assuming a known baseline hazard shift, a correctly specified imputation model, and error-free external predictions.
} \label{tab:simul 2_Gamma}

\vspace{1em}
\begin{tabular}{l rrr @{\hskip 6pt} rrr @{\hskip 6pt} rrr}
\hline
 & \multicolumn{3}{c}{EPI} & \multicolumn{3}{c}{EPI-trueZ} & \multicolumn{3}{c}{EPI-oracle}  \\
\cline{2-4} \cline{5-7}  \cline{8-10} 
             & bias & MSE & CP & bias & MSE & CP & bias & MSE & CP \\
\hline 
$\gamma_0$   &  0.006 & 0.003 & 0.983 &  0.001 & 0.001 & 0.950 &  0.003 & 0.001 & 0.960  \\
$\gamma_1$   & -0.018 & 0.013 & 0.963 & -0.004 & 0.013 & 0.960 & -0.002 & 0.013 & 0.957  \\
$\gamma_2$   & -0.018 & 0.013 & 0.960 & -0.002 & 0.012 & 0.957 & -0.004 & 0.012 & 0.960  \\
$\gamma_3$   & -0.004 & 0.094 & 0.993 &   &  &  &        &       &        \\
$\gamma_4$   &  0.014 & 0.116 & 0.980 &   &  &  &        &       &        \\
$\gamma_5$   & -0.053 & 0.104 & 0.990 &   &  &  &        &       &        \\
$\gamma_6$   & -0.002 & 0.003 & 0.980 &   &  &  &        &       &        \\
$\gamma_7$   & -0.003 & 0.003 & 0.990 &   &  &  &        &       &        \\
$\sigma^2$   &  0.010 & 0.001 & 0.910 &  0.020 & 0.001 & 0.903 &  0.019 & 0.006 & 0.910  \\
$\nu$        &  0.080 & 0.032 & 0.937 &  0.081 & 0.034 & 0.933 &  - & - &   \\
\hline 
\end{tabular}
\end{table}

\vspace{5em}
\section{Real Data Analysis}

\subsection{Description of the Clinical Trials on the Prostate Cancer Data}

\newcolumntype{Y}{>{\raggedright\arraybackslash}X}
\begin{sidewaystable}[!p]
\centering
\caption{Data description of the clinical trials on metastatic Castration-Resistant Prostate Cancer patients: The COU-AA-302 is the internal dataset of our study. The CALGB 90401 and PREVAIL are the model development cohorts of Halabi's and PREVAIL calculators, respectively. Note that the primary endpoint of the CALGB 90401 trial is the time when the study achieved the target 748 deaths,and group-specific mortalities are not available. } \label{tab:data description}
\scriptsize 
\renewcommand{\arraystretch}{0.7}

\vspace{1em}
\begin{tabularx}{\textwidth}{lYYY}
\hline
& Internal data & Halabi's cohort & PREVAIL cohort \\
\hline 

Enrollment period
& Apr 2009 -- Jun 2010 
& May 2005 -- Dec 2007 
& Sep 2010 -- Sep 2012  \\ 
\hline 

Sample size (treated/control)
& 1,088 (546/542) & 1,050 (524/526) & 1,717 (872/845) \\
\hline  

Drug
& Prednisone with or without abiraterone acetate
& Docetaxel and Prednisone with or without Bevacizumab
& Enzalutamide or placebo \\
\hline 

\multirow{4}{*}{Inclusion criteria}
& ECOG 0 or 1 & ECOG 0--2 & ECOG 0 or 1  \\

& Testosterone $<50$ ng/dL
& Testosterone $<50$ ng/dL 
& Testosterone $<50$ ng/dL \\

& Pain score 0--3 &  & Pain score 0--3 \\

& Continued androgen-deprivation therapy 
&  
& Continued androgen-deprivation therapy \\
\hline 

\multirow{3}{*}{Exclusion criteria}
& Visceral metastases
& Clinically significant cardiac disease or neuropathy grade $\ge 2$
& History of seizure or predisposition to seizure \\

& Previous ketoconazole chemotherapy 
& Previous cytotoxic chemotherapy or antiangiogenic agents 
& Previous cytotoxic or ketoconazole chemotherapy \\

&  & Brain metastases &  \\
\hline

Follow-up range
& Day 1 to 60 months
& Day 1 to 60 months
& Day 1 to 36 months \\
\hline

Overall mortality
& 354 (65\%) in the treated group vs. 387 (71\%) in control group 
& Overall 748 deaths
& 241 (28\%) in the treated group vs. 299 (35\%) in the control group \\
\hline

Median survival (months)
& 34.7 in the treated group vs. 30.3 in the control group 
& 22.6 in the treated group vs. 21.5 in the control group
& 32.4 in the treated group vs. 30.2 in the control group \\
\hline

Risk factors
& PSA, LDH$\geq$ULN, albumin, hemoglobin, bone metastases, visceral metastases, lymph node metastases, ECOG, NLR$<$2, pain score $\geq$ 2, time from diagnosis to randomization, decipher score
& PSA, LDH$\geq$ULN, albumin, hemoglobin, bone metastases, visceral metastases, lymph node metastases, ECOG, opioid use 
& PSA, LDH$\geq$ULN, albumin, hemoglobin, bone metastases, visceral metastases, NLR$<$2, pain score $\geq$ 2, time from diagnosis to randomization \\
\hline

Survival prediction time
& --
& 18, 24, 30, 36, and 48-month
& 1, 2, 3, and 5-year \\
\hline
\end{tabularx}
\vspace{1em}

\parbox{\linewidth}{\scriptsize
Abbreviations: ECOG, Eastern Cooperative Oncology Group performance status; PSA, prostate-specific antigen; LDH, lactate dehydrogenase; ULN, upper limit of normal; ALP, alkaline phosphatase; NLR, neutrophil to lymphocyte ratio; \\
}
\end{sidewaystable}

\clearpage 
\subsection{Real Data Analysis Results}

\begin{table}[h!]
    \centering
    \caption{Estimated regression coefficients (Est.) and standard errors (SE) of working imputation models under the fully extended external prediction informed Cox (EPI) model for the prostate cancer study combining Halabi's and PREVAIL predictions. Variance indicates the residual error terms of the linear models for continuous variables (Time from diagnosis and Decipher score). Columns correspond to the non-overlapping covariates in the two prediction models, accordingly.} \label{tab:realdata working coef}

    \vspace{1em}
    \begin{tabular}{l c cc @{\hskip 6pt} ccc}
    \hline 
        & ECOG  & Pain score  & Time from diagnosis & Decipher score  \\
    \cline{2-3} \cline{4-5} 
    Variables             & Est. (SE) & Est. (SE) & Est. (SE) & Est. (SE) \\
    \hline
    Intercept             & -7.177 (4.413) &  1.089 (3.847) & -3.257 (7.264) & -0.016 (0.965) \\
    log(PSA)              &  0.167 (0.189) &  0.108 (0.176) &  0.131 (0.301) &  0.130 (0.149) \\
    LDH$\geq$ULN          &  0.042 (0.870) &  0.380 (0.775) &  1.544 (1.408) &  0.773 (0.632) \\
    Albumin               &  1.046 (0.929) & -1.139 (0.888) & -0.420 (1.497) & -0.171 (0.442) \\
    Hemoglobin            &  0.033 (0.260) &  0.077 (0.240) &  0.772 (0.446) &  0.005 (0.144) \\
    ALP$\geq$ULN          & -0.766 (0.802) &  0.444 (0.622) &  0.972 (1.176) &  0.280 (0.471) \\
    Bone metastases       &  0.535 (0.687) &  1.127 (0.757) & -0.596 (1.076) &  0.207 (0.429) \\
    ECOG                  &                &  0.238 (0.624) &  0.125 (1.083) & -0.335 (0.452) \\
    Pain score$\geq$2     &                &                &  0.462 (1.041) &  1.163 (0.540) \\
    Time from diagnosis   &                &                &                &  0.118 (0.055) \\
    \hline
    Variance             &                &                & 19.253 (2.948) &  5.243 (0.787) \\
    \hline
    \end{tabular}
    \vspace{1em}
    
    \parbox{1.0\linewidth}
    {\footnotesize
    Abbreviations: PSA, prostate-specific antigen; LDH, lactate dehydrogenase; ULN, upper limit of normal; ALP, alkaline phosphatase; ECOG, Eastern Cooperative Oncology Group performance status. \\
    }
\end{table}

\begin{table}[h!]
\centering
\caption{Estimated regression coefficients (Est.) and standard errors (SE) of the Cox models for the prostate cancer study combining Halabi's predictions. EPI is the fully extended external prediction informed Cox model that allows heterogeneous baseline hazard, sampling errors in the external predictions, and a working model for $Z$. $\nu^H$ is the parameter accounting for the baseline hazard shift of Halabi's model. A column labeled Halabi reports the regression coefficient estimates copied from the original study.} \label{tab:realdata2}

\vspace{1em}
    \begin{tabular}{l @{\hskip 6pt} ccc} 
    \hline
     & Internal-only & EPI & Halabi  \\
    \cline{2-4}
    Variables             & Est. (SE) & Est. (SE) & Est. (SE) \\
    \hline
    log(PSA)              &  0.168 (0.096) &  0.080 (0.067) & {\color{black} 0.009 (0.006)} \\
    LDH$\geq$ULN          & -0.267 (0.467) &  0.030 (0.301) & {\color{black} 0.146 (0.039)} \\
    Albumin               &  0.124 (0.430) & -0.004 (0.262) & {\color{black}-0.051 (0.029)} \\
    Hemoglobin            & -0.013 (0.120) & -0.009 (0.073) & {\color{black}-0.027 (0.014)} \\
    ALP$\geq$ULN          &  0.404 (0.352) &  0.237 (0.236) & {\color{black} 0.064 (0.029)} \\
    Bone metastases       &  0.494 (0.361) &  0.453 (0.313) & {\color{black}  -- } \\
    ECOG                  & -0.863 (0.382) & -0.345 (0.240) & {\color{black}-0.134 (0.035)} \\
    Decipher score        &  0.033 (0.062) &  0.037 (0.058) &    \\
    Treatment group       & -0.062 (0.294) & -0.065 (0.271) &    \\
    \hline
    $\nu^H$  &  & 0.177 (0.218) &  \\
    \hline
    \end{tabular}
\vspace{1em}

\parbox{0.95\linewidth}
{\footnotesize
Abbreviations: PSA, prostate-specific antigen; ULN, upper limit of normal; ALP, alkaline phosphatase; ECOG, Eastern Cooperative Oncology Group performance status. \\
}
\end{table}

\begin{table}[h!]
    \centering
    \caption{Estimated regression coefficients (Est.) and standard errors (SE) of the working imputation model under the fully extended external prediction informed Cox (EPI) model for the prostate cancer study combining Halabi's predictions. Variance indicates the residual error term of the linear model for the decipher score.} \label{tab:realdata2 working coef}

    \vspace{1em}
    \begin{tabular}{l c }
    \hline
                          & Decipher score  \\
    \cline{2-2}
    Variables             &  Est. (SE) \\
    \hline
    Intercept             &  0.043 (0.984) \\
    log(PSA)              &  0.170 (0.151) \\
    LDH$\geq$ULN          &  0.967 (0.676) \\
    Albumin               & -0.261 (0.489) \\
    Hemoglobin            &  0.083 (0.151) \\
    ALP$\geq$ULN          &  0.429 (0.506) \\
    Bone metastases       &  0.277 (0.465) \\
    ECOG                  & -0.289 (0.465) \\
    \hline
    Variance              &  5.931 (0.886) \\
    \hline
    \end{tabular}
    \vspace{1em}
    
    \parbox{1.0\linewidth}
    {\footnotesize
    Abbreviations: PSA, prostate-specific antigen; LDH, lactate dehydrogenase; ULN, upper limit of normal; ALP, alkaline phosphatase; ECOG, Eastern Cooperative Oncology Group performance status.}
\end{table}

\clearpage
\bibliography{A_reference}

\label{lastpage}